\documentclass[12pt,draftclsnofoot,onecolumn]{IEEEtran}
\usepackage{amsmath,amssymb,amsfonts,mathrsfs,bm}
\usepackage{amstext}
\usepackage{color}
\usepackage{upgreek}
\usepackage{multicol}
\usepackage{algorithm}
\usepackage{algorithmic}
\usepackage{graphicx}
\usepackage{paralist}
\usepackage[numbers,sort&compress]{natbib}
\usepackage{booktabs}
\usepackage{multirow}
\usepackage{caption}
\usepackage{diagbox}
\usepackage{subfigure}
\usepackage{booktabs} 

\IEEEoverridecommandlockouts

\begin{document}

\title{Multicell Power Control under Rate Constraints with Deep Learning}
\author{
\IEEEauthorblockN{Yinghan Li, Shengqian Han, \textit{Member, IEEE}, Chenyang Yang, \textit{Senior}\\ \textit{Member, IEEE}}
\thanks{
The authors are with the School of Electronics and Information Engineering, Beihang University, Beijing 100191, China (e-mail:\{llyyhh, sqhan, cyyang\}@buaa.edu.cn).} 
}

\maketitle

\begin{abstract}
In the paper we study a deep learning based method to solve the multicell power control problem for sum rate maximization subject to per-user rate constraints and per-base station (BS) power constraints. The core difficulty of this problem is how to ensure that the learned power control results by the deep neural network (DNN) satisfy the per-user rate constraints. To tackle the difficulty, we propose to cascade a projection block after a traditional DNN, which projects the infeasible power control results onto the constraint set. The projection block is designed based on a geometrical interpretation of the constraints, which is of low complexity, meeting the real-time requirement of online applications. Explicit-form expression of the backpropagated gradient is derived for the proposed projection block, with which the DNN can be trained to directly maximize the sum rate via unsupervised learning. We also develop a heuristic implementation of the projection block to reduce the size of DNN. Simulation results demonstrate the advantages of the proposed method over existing deep learning and numerical optimization~methods, and show the robustness of the proposed method with the model mismatch between training and testing~datasets.
\end{abstract}



\section{Introduction}
\label{s:1}


Multicell power control for sum rate maximization (SRM) is a well-known non-convex problem in the wireless communication community. Various efforts have been devoted to find efficient solutions to this problem.
Early works studied different approximations to the problem, e.g.,~\cite{Chiang2007Power,Papandriopoulos2009SCALE}.
For high signal-to-interference plus noise ratio (SINR), the SRM problem was approximated as a geometric
programming problem by ignoring the noise~\cite{Chiang2007Power}. For medium to low SINR, the SRM problem was convexified by using the monomial approximation for posynomial~\cite{Chiang2007Power} or the successive logarithmic approximations~\cite{Papandriopoulos2009SCALE}.
In~\cite{Shi2011}, an iterative algorithm, namely weighted sum mean-square error minimization (WMMSE), based on the equivalence between sum rate and sum MSE was proposed for the optimization of multicell beamformer. It is directly applicable to the SRM problem in single-antenna systems, and guarantees the convergence to a
stationary point~\cite{Shi2011}.

With the prevalence of deep learning, the deep neural network (DNN) based approaches have been recently studied to solve the SRM problem. In~\cite{Sun2018}, supervised learning was employed to maximize the sum rate subject to per-base station (BS) transmit power constraints, where the DNN was trained by taking the power control results obtained by WMMSE~\cite{Shi2011} as the ground truth (i.e., label). This method provides a low-complexity implementation of WMMSE via DNN, but its performance cannot exceed WMMSE. Some works studied unsupervised learning methods by directly using the sum rate as the loss function for the training of DNN, e.g.,~\cite{Liang2018Towards,Lee2019a}. In~\cite{Liang2018Towards}, the ensemble learning was employed on the basis of unsupervised learning to enhance the performance, which can outperform WMMSE in the scenarios with high SINR or few users. In~\cite{Meng2019}, deep reinforcement learning was employed to solve the  SRM problem, where the power of a BS was discretized into multiple levels as the actions for selection.


A challenge of applying deep learning to solve the SRM problem is how to deal with the constraints. Some simple constraints can be easily handled by choosing proper activation function for the output layer. In~\cite{Sun2018,Liang2018Towards,Lee2019a}, for instance, the per-BS power constraints were considered. To make the learned power control results satisfy the constraints, the activation function for the output layer was selected as a truncated $\operatorname{ReLU}$ function in~\cite{Sun2018}, the $\operatorname{Sigmoid}$ function in~\cite{Liang2018Towards}, and a modified $\operatorname{Tanh}$ function in~\cite{Lee2019a}.
For general constraints, e.g., the per-user rate constraints, however, finding an appropriate activation function for the output layer is non-trivial. To tackle the difficulty, prior works proposed to incorporate the penalty of the constraint violation into the loss function, which was supposed to incentivize the DNN output to meet the constraints~\cite{Liang2018Towards,Lee2020}. In particular, the penalty was added to the loss function in~\cite{Liang2018Towards} while was multiplied to the loss function in~\cite{Lee2020}. The performance of this kind of methods depends on the selection of the penalty factor: a small penalty factor cannot ensure the constraints to be satisfied, while a large penalty factor distorts the objective function and degrades the performance.
Another deep learning framework for the constrained optimization problem was proposed based on the duality theory by~\cite{Eisen2019,Sun2019}. It transforms the constrained problem into an unconstrained problem by using the Lagrangian approach, with which a DNN is trained to parameterize both the primal and dual variables. However, due to the residual error of DNN for parameterizing the dual variables, the methods cannot guarantee that the constraints are always satisfied.

In this paper, we propose a deep learning solution to the SRM problem, which always guarantees the per-user rate constraints. The main contributions of the paper are summarized as~follows.\footnote{The codes for the proposed methods in the paper are made available to promote reproducible research: https://github.com/Leeyyhh/SRnet-and-SRNet-Heu-for-power-control.}
\begin{itemize}
  \item We propose a novel DNN structure by cascading a projection block after a traditional DNN, where the projection block ensures the per-user constraints to be satisfied. Different from the classic $\ell_2$ projection that requires iterative computations as studied in \cite{Amos17,Lee2019}, the proposed projection block is designed based on a geometrical interpretation of the constraints, which is of low complexity and meets the real-time requirement of online applications.
  \item We derive the explicit-form expression of the backpropagated gradient for the proposed projection block, which enables an unsupervised learning of the power control to directly maximize the sum rate. We also develop a heuristic method to simplify the projection block, which is able to reduce the network size of the DNN. 
  \item Simulation results demonstrate the performance advantages of the proposed method compared to existing deep learning and typical numerical optimization methods, and show that the proposed method is not sensitive to the model mismatch between training and testing datasets, which makes it attractive for practical applications.

\end{itemize}

\section{System Model}
\label{s:2}
Consider the downlink transmission of $K$ cells, where each cell serves a single user equipment (UE).
The received signal by the UE in the $i$-th cell, denoted by UE$_i$, is expressed as
\begin{align}
  y_i = \sqrt{\alpha_{ii}}h_{ii}\sqrt{p_i}s_i + \sum_{j=1, j\neq i}^K \sqrt{\alpha_{ij}}h_{ij}\sqrt{p_j}s_j + w_i,
\end{align}
where the BSs and the UEs are equipped with a single antenna, $s_i\sim\mathcal{CN}(0,1)$ is the transmit signal of the $i$-th BS (BS$_i$) to UE$_i$, $s_i$ and $s_j$ are independent for $i\neq j$, $\alpha_{ij}$ and $h_{ij}$ are the large- and small-scale channels from BS$_j$ to UE$_i$, $p_{j}$ is the transmit power of BS$_j$, and $w_i\sim\mathcal{CN}(0,\sigma^{2})$ is the additive white Gaussian noise at UE$_i$.

The SINR of UE$_i$, denoted by $\gamma_{i}$, can be expressed as
\begin{equation} \label{E:SINR}
\gamma_{i}=\frac{\alpha_{ii}\left|h_{ii}\right|^{2}p_{i}}{\sum_{j=1, j \neq i}^{K}\alpha_{ij}\left|h_{ij}\right|^{2}p_{j}+\sigma^{2}}.
\end{equation}

The SRM problem subject to per-user rate constraints and per-BS power constraints can be formulated as
\begin{subequations} \label{E:problem}
    \begin{align}
    \max_{\left\{p_{i}\right\}}\  & \sum_{i=1}^{K} \log \left(1+\gamma_{i}\right) \label{E:problem1-1} \\
    s.t.\  &  \log \left(1+\gamma_{i}\right) \geq r_{i,\text{min}},\ \forall i, \label{E:problem-2} \\
           & 0 \leq p_{i} \leq P_{\text{max}},\ \forall i, \label{E:problem-3}
    \end{align}
\end{subequations}
where $r_{i,\text{min}}$ is the minimum rate required by UE$_i$, and $P_{\text{max}}$ is the maximal transmit power of each BS.

Constraint \eqref{E:problem-2} can be rewritten as $\gamma_i\geq2^{r_{i,\text{min}}}-1\triangleq \gamma_{i,\text{min}}$, $\forall i$, where $\gamma_{i,\text{min}}$ stands for the minimum SINR required by UE$_i$.
With the expression of $\gamma_i$ in \eqref{E:SINR}, the constraint $\gamma_i\geq\gamma_{i,\text{min}}$ can be expressed as a linear constraint, so that problem \eqref{E:problem} can be rewritten as
\begin{subequations} \label{E:problem-B}
    \begin{align}
    \max_{\mathbf{p}}\  & \sum_{i=1}^{K} \log \left(1+\gamma_{i}\right) \label{E:problem1-B1} \\
    s.t.\  &  \mathbf{B}\mathbf{p} \succeq \mathbf{q}, \label{E:problem-B2} \\
           & \mathbf{0} \preceq \mathbf{p} \preceq P_{\text{max}}\mathbf{1}, \label{E:problem-B3}
    \end{align}
\end{subequations}
where matrix $\mathbf{B}\in\mathbb{C}^{K\times K}$ is defined as $[\mathbf{B}]_{ij} = -\gamma_{i,\text{min}}\alpha_{ij}|h_{ij}|^2, \forall j\neq i$, and $[\mathbf{B}]_{ii} = \alpha_{ii}|h_{ii}|^2, \forall i$, vector $\mathbf{q}=\left[q_1,\dots,q_K\right]^T\in\mathbb{C}^{K\times 1}$ is defined as $q_i = \gamma_{i,\text{min}}\sigma^2$, $\mathbf{p} = [p_1, \dots, p_K]^T$, $[\mathbf{X}]_{ij}$ denotes the element of matrix $\mathbf{X}$ on the $i$-th row and $j$-th column, $\preceq$ and $\succeq$ indicate element-wise inequalities, and $\mathbf{0}$ and $\mathbf{1}$ denote all-zero and all-one vectors.

Problem \eqref{E:problem-B} is non-convex since the objective function \eqref{E:problem1-B1} is non-convex, making it difficult to find the global optimal solution. To obtain a low-complexity solution to the problem for online applications, we follow the idea of ``learning to optimize''~\cite{Sun2018} and resort to deep learning to parameterize the solution. Although the optimality of the learned solution has no theoretical guarantee, numerous works have shown that the solutions of deep learning usually perform reasonably well.

\section{Power Control by DNN}
\label{s:3}
\begin{figure}
	\centering
	\includegraphics[width=0.7\textwidth]{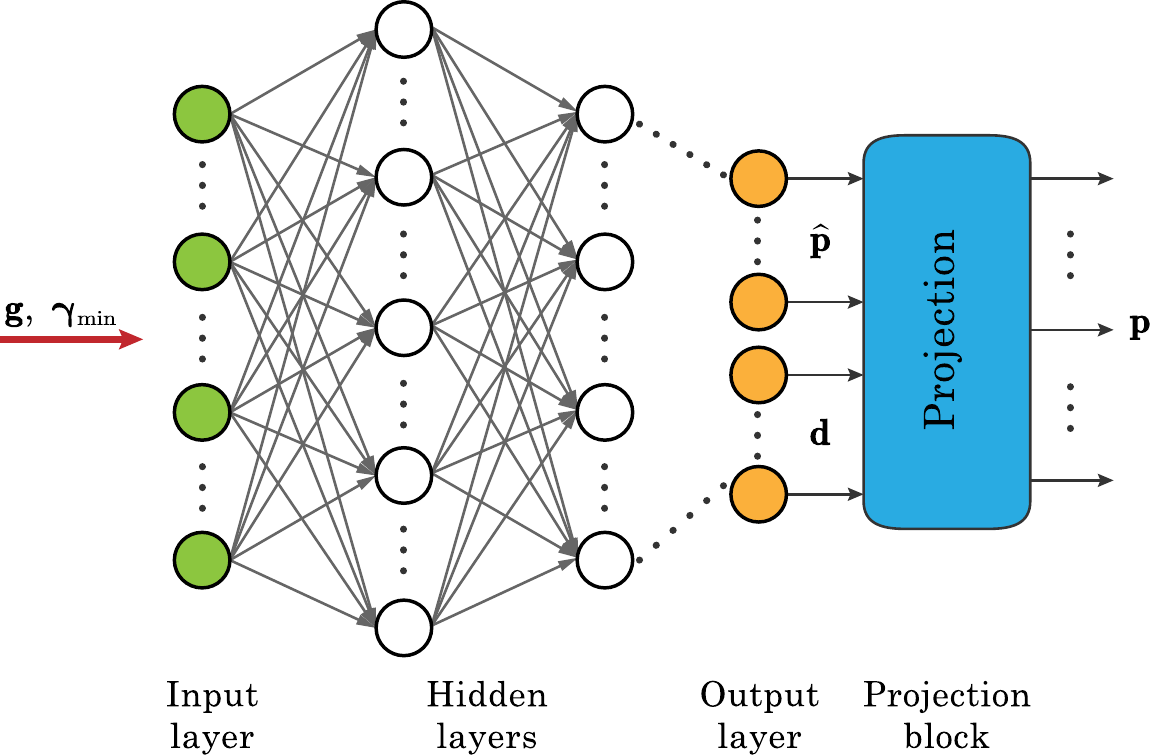}
     \caption{Structure of the employed DNN.}
	\label{fig:network}
\end{figure}

The difficulty of using DNN to solve the considered SRM problem lies in how to regulate the network output to satisfy the per-user rate constraints. Existing activation functions are mostly defined as scalar-to-scalar functions,\footnote{The $\operatorname{Softmax}$ function is an exception that is a vector-to-vector function. It can be used to constrain the sum of multiple variables to be one, but is not applicable to the rate constraints of multiple UEs.} which can effectively model the constraints for each individual variable but is unable to address the constraints that are coupled with multiple variables.
To tackle the difficulty, we propose to cascade a projection block after the output layer of a traditional DNN, which projects an infeasible power generated by the output layer onto the constraint set. The structure of the employed network is illustrated in Fig~\ref{fig:network}. The design of the projection block faces \textbf{two challenges}: 1) the projection should be of low complexity and suitable for real-time implementation; and 2) the backpropagated gradients of the projection block should be available so that the end-to-end gradient backpropagation is allowed for network training.
As we will see later, the widely used $\ell_2$ projection, i.e., finding a point in the constraint set that has the minimal distance from an infeasible power, cannot satisfy the above two conditions.

In the following, we first elaborate the employed DNN, and then focus on the design of the projection block.

\subsection{Design of DNN}
As shown in Fig~\ref{fig:network}, the employed DNN consists of an input layer, multiple hidden layers, an output layer, and a projection block, where the former three parts have no difference from a traditional fully-connected DNN.

The input layer has $K^{2}\!+\!K$ nodes, which corresponds to the input $\mathbf{x} = [\mathbf{g}^T,  \boldsymbol{\gamma}_{\text{min}}^T]^T$, where $\mathbf{g} = [\alpha_{11}|h_{11}|^2, \dots, \alpha_{ij}|h_{ij}|^2, \dots, \alpha_{KK}|h_{KK}|^2]^T$ collects the large- and small-scale channel gains from $K$ BSs to $K$ UEs, and $\boldsymbol{\gamma}_{\text{min}}=[\gamma_{1,\text{min}},\dots,\gamma_{K,\text{min}}]^T$ collects the minimum SINR required by $K$ UEs.
The hidden layers are fully-connected between the input and output layers.
The output layer has $2K$ nodes. The first $K$ nodes can be understood as a temporary power control result for $K$ BSs, denoted by $\hat{\mathbf{p}}= [\hat{p}_1, \dots, \hat{p}_K]^T$, which may not satisfy the per-user rate constraints. The other $K$ nodes correspond to a group of the so-called ``distances'', denoted by ${\mathbf{d}}= [d_1, \dots, d_K]^T$, which are used in the design of the projection block and will be detailed later.
The projection block takes $\hat{\mathbf{p}}$ and ${\mathbf{d}}$ as input and outputs the final power control result $\mathbf{p}$, which has the functionality of projecting $\hat{\mathbf{p}}$ onto the set defined by per-user rate constraints.

Due to lack of the optimal solution to the SRM problem as training labels, supervised learning is not applicable to the problem. This motivates us to train the DNN in an unsupervised learning manner to directly maximize the sum rate, which is called \emph{SRNet} in the sequel for simplicity. The loss function to be minimized for training is defined as
\begin{equation} \label{E:loss}
    J_{\operatorname{SRNet}}(\boldsymbol{\theta})=-\frac{1}{M_{tr}}\sum_{m=1}^{M_{tr}} \sum_{i=1}^{K} \log \left(1+\gamma_{i}\left(\mathbf{x}^{(m)},\boldsymbol{\theta}\right)\right) \triangleq \frac{1}{M_{tr}}\sum_{m=1}^{M_{tr}} J^{(m)}(\boldsymbol{\theta}),
\end{equation}
where $\boldsymbol{\theta}$ denotes the set of trainable network parameters, $\gamma_{i}(\mathbf{x}^{(m)}, \boldsymbol{\theta})$ is the SINR achieved by the learned power control under a specific input realization $\mathbf{x}^{(m)}$ and network parameters $\boldsymbol{\theta}$, $M_{tr}$ is the number of training samples, and $J^{(m)}(\boldsymbol{\theta})\triangleq-\sum_{i=1}^{K}$ $\log \left(1+\gamma_{i}\left(\mathbf{x}^{(m)},\boldsymbol{\theta}\right)\right)$.
It can be found that the loss function \eqref{E:loss} is minimized when the sum rate for every input realization $\mathbf{x}^{(m)}$, i.e., $-J^{(m)}(\boldsymbol{\theta})$, is maximized by the trained parameters $\boldsymbol{\theta}$.


\subsection{Basic Principle of Designing Projection Block}
\label{s:projection}

To project an infeasible $\hat{\mathbf{p}}$ generated by the output layer onto the constraint set, the $\ell_2$ projection is a widely used approach. It finds a feasible point $\mathbf{p}$ in the constraint set to minimize the distance from $\hat{\mathbf{p}}$ by solving the following problem~\cite{Boyd04}
\begin{subequations} \label{E:proj}
  \begin{align}
    \min_{\mathbf{p}}\  & ||\mathbf{p} - \hat{\mathbf{p}}||^2 \label{E:proj-problem1-1} \\
    s.t.\  &  \mathbf{B}\mathbf{p} \succeq \mathbf{q}, \label{E:proj-B2} \\
           &  \mathbf{p} \preceq P_{\text{max}}\mathbf{1}, \label{E:proj-B3}
  \end{align}
\end{subequations}
where $||\cdot||$ is the $\ell_2$ norm.

By comparing \eqref{E:problem-B3} and \eqref{E:proj-B3}, one can find that we have omitted the constraint $\mathbf{p} \succeq \mathbf{0}$ in \eqref{E:proj-B3}. We will see soon that omitting this constraint does not affect the optimality of the projection because $\mathbf{p} \succeq \mathbf{0}$ always holds for a feasible problem, i.e., the minimum rate constraints \eqref{E:proj-B2} induce $\mathbf{p} \succeq \mathbf{0}$. Meanwhile, omitting the constraint can facilitate the subsequent design of the projection block.

The per-user rate or SINR constraints in \eqref{E:problem-2} or \eqref{E:problem-B2} are not always feasible under the multicell interfering environment. A criteria to judge the feasibility was given by \cite[Theorem 2.2]{Chiang2008}, which is summarized as follows.
\begin{itemize}
  \item Defining $\mathbf{p}_0 = \mathbf{B}^{-1}\mathbf{q}$, then problem \eqref{E:problem} or \eqref{E:problem-B} is feasible if and only if $\mathbf{0} \preceq \mathbf{p}_0 \preceq P_{\text{max}}\mathbf{1}$~holds.
  \item Any feasible power $\mathbf{p}$ to problem \eqref{E:problem} or \eqref{E:problem-B} satisfies that $\mathbf{p}~\succeq~\mathbf{p}_0$.
\end{itemize}

With this criteria, we can readily judge the feasibility of problem \eqref{E:problem} or \eqref{E:problem-B}, i.e., whether the minimum rate constraints of UEs are achievable or not. Moreover, it is easy to find that $\mathbf{p} \succeq \mathbf{0}$ always holds for a feasible problem, which thus can be omitted in \eqref{E:proj}. In the sequel, we suppose that the per-user rate constraints are feasible.
\begin{figure}
	\centering
	\includegraphics[width=0.75\textwidth]{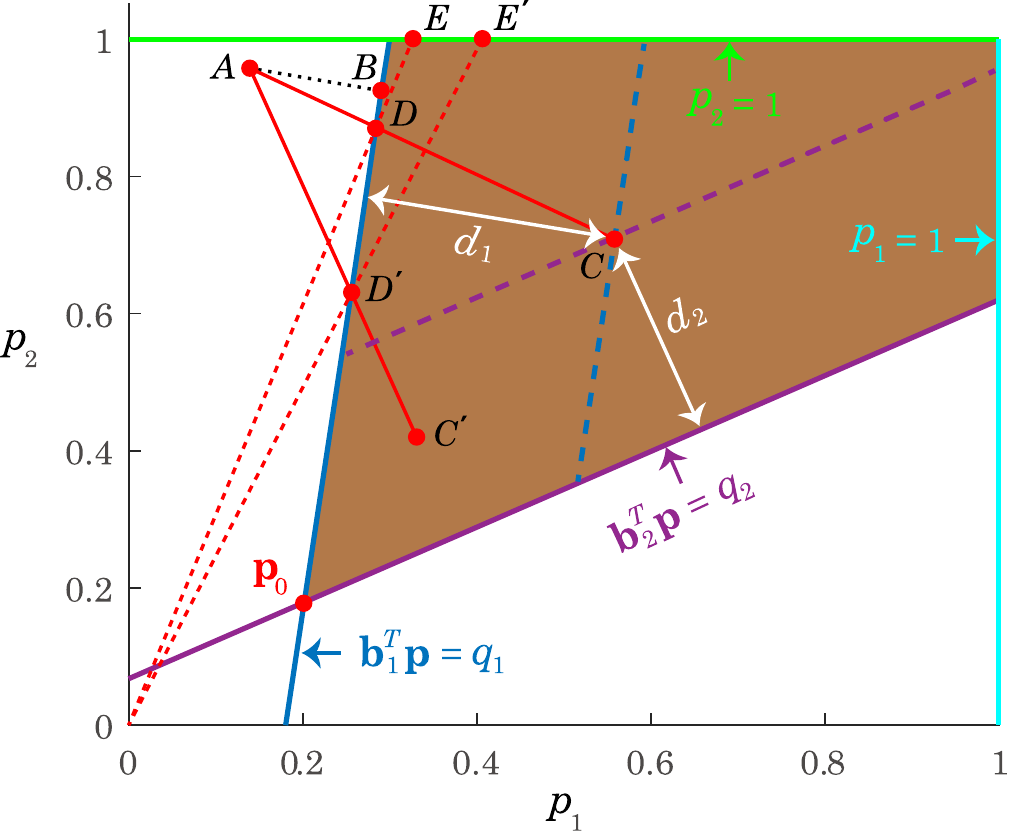}
    \caption{Illustration of the proposed projection method.}
	\label{fig:projection}
\vspace{-0.2cm}
\end{figure}

In Fig.~\ref{fig:projection}, we provide a geometrical demonstration of the projection process, where $K=2$ cells and $P_{\text{max}}=1$ are considered.
The shaded area is the feasible region of $\mathbf{p}$. Let point $A$ denote the infeasible power $\hat{\mathbf{p}}$. Then, point $B$, having the shortest distance from $A$ in the feasible region, is the optimal solution to problem \eqref{E:proj}, i.e., the $\ell_2$ projection of $A$. However, finding point $B$ by solving the projection problem \eqref{E:proj} requires an iterative algorithm despite the problem is convex, which is unsuitable for real-time application. Moreover, computing the backpropagated gradients from point $B$ to point $A$ is very complicated. For instance, the method proposed by~\cite{Amos17} was applied in~\cite{Lee2019} to compute the backpropagated gradients, which requires to find the optimal primal and dual variables of problem \eqref{E:proj}. In summary, directly using the $\ell_2$ projection cannot overcome the aforementioned two challenges at the beginning of this section.

In the following, we aim to find a novel projection of point $A$ to overcome the two challenges. The goal is to find an approximation to the projection point $B$, which is expected to have low complexity and meanwhile facilitate the network training. The main procedures to obtain such an approximation are summarized below, as illustrated by Fig.~\ref{fig:projection}.
\begin{itemize}
  \item[1)] Find an interior point in the feasible region, denoted by point $C$, corresponding to the power $\mathbf{p}^C = [p_1^C, \dots, p_K^C]^T$.
  \item[2)] Draw the segment $AC$ by connecting point $C$ and $A$.
  \item[3)] Obtain the cross point between the segment $AC$ and the boundary of the feasible region, denoted by point $D$, corresponding to the power $\mathbf{p}^D = [p_1^D, \dots, p_K^D]^T$, which can be interpreted as an approximation to the optimal projection point $B$.
  \item[4)] Obtain the output of the projection block, denoted by point $E$, whose power equals to $\mathbf{p}^E=\frac{P_{\text{max}}}{\max\{\mathbf{p}^D\}}\mathbf{p}^D$. This step follows from the observation from \eqref{E:SINR} that equally scaling up the powers of all BSs increases the~SINR. Thus, it is ensured that the SINR achieved by point $E$ is not lower than that by point $D$. Geometrically, point $E$ is the cross point between the ray connecting the original point (with $\mathbf{p}=\mathbf{0}$) and $D$ and the boundary of the feasible~region.
\end{itemize}

In the above procedures, we suppose that point $A$ is out of the feasible region. If point $A$ is in the feasible region, i.e., $\hat{\mathbf{p}}$ satisfies constraints \eqref{E:problem-B2} and \eqref{E:problem-B3}, then the projection is no longer necessary and point $D$ is just equal to point $A$. We only need to scale $\mathbf{p}^D$ as step 4) to obtain point $E$.

We next elaborate the approaches of finding point $C$ and $D$, respectively.

\subsection{Finding Point $C$}
\subsubsection{\underline{Optimizing Point $C$}}
The location of the interior point $C$ has a direct impact on the projection result. As shown in Fig.~\ref{fig:projection}, if we replace point $C$ with point $C'$, then the obtained projection will become point $D'$, which is clearly farther than point $D$ from the optimal $\ell_2$ projection point~$B$.

To optimize point $C$, we first examine how to express an interior point in the feasible region. Let $\mathbf{b}_i^T$ denote the $i$-th row of $\mathbf{B}$, and $d_i$ denote the distance from the interior point $C$ to the hyperplane $\mathbf{b}_i^T\mathbf{p}=q_i$, $i=1,\dots,K$. Then, we can obtain a hyperplane that is in parallel with the hyperplane $\mathbf{b}_i^T\mathbf{p}=q_i$ and meanwhile contains point $C$, as illustrated with dash lines in Fig.~\ref{fig:projection}, which can be expressed~as
\begin{align}
\label{E:parallel}
  \mathbf{b}_i^T\mathbf{p} = q_i + \|\mathbf{b}_i\|_2d_i, \ i=1,\dots,K.
\end{align}

Then, the interior point $C$, i.e., the cross point of the $K$ hyperplanes given by \eqref{E:parallel}, satisfies the equations
\begin{align}
  \mathbf{B}\mathbf{p} = \mathbf{q} + \mathrm{diag}(\mathbf{B}\mathbf{B}^T)^{\frac{1}{2}}\mathbf{d},
\end{align}
from which the power corresponding to point $C$ can be solved as
\begin{align} \label{E:central_point}
  \mathbf{p}^C(\mathbf{d}) = \mathbf{p}_0 + \mathbf{B}^{-1}\mathrm{diag}(\mathbf{B}\mathbf{B}^T)^{\frac{1}{2}}\mathbf{d},
\end{align}
where $\mathbf{d} = [d_1,\dots,d_K]^T$, $\mathrm{diag}(\mathbf{X})$ denotes a diagonal matrix consisting of the diagonal element of matrix $\mathbf{X}$, and $\mathbf{p}_0 = \mathbf{B}^{-1}\mathbf{q}$ as defined before.

As a result, to optimize point $C$ is equivalent to optimize the distance $\mathbf{d}$, which should ensure that the resultant point $C$ is located within the feasible region, i.e., satisfying constraints \eqref{E:proj-B2} and \eqref{E:proj-B3}. From \eqref{E:parallel}, we know that constraint \eqref{E:proj-B2} is satisfied as long as $d_i\geq 0$, $\forall i$. Constraint \eqref{E:proj-B3} requires
\begin{align} \label{E:central_point-d}
  \mathbf{p}^C(\mathbf{d}) \preceq P_{\text{max}}\mathbf{1}.
\end{align}

Next, we focus on the optimization of $\mathbf{d}$ subject to \eqref{E:central_point-d} and $d_i\geq 0$, $\forall i$.

\subsubsection{\underline{Learning Distance $\mathbf{d}$}}
It is desirable to find the distance $\mathbf{d}$ that makes the output of the projection block be able to maximize the sum rate. This, however, is difficult due to the non-convexity of the sum rate. Thus, we resort to use DNN to learn $\mathbf{d}$, which corresponds to the second-half nodes of the output layer in Fig.~\ref{fig:network}.

The difficulty of learning $\mathbf{d}$ lies in the selection of the activation function for the corresponding nodes in the output layer, which should be able to ensure that \eqref{E:central_point-d} is satisfied. This is challenging because \eqref{E:central_point-d} makes the elements of $\mathbf{d}$ coupled. To circumvent the difficulty, we propose to optimize $\mathbf{d}$ in a subset rather than the whole set of the feasible region of $\mathbf{d}$. The subset is defined as $\{\mathbf{d}|0\leq d_i\leq d_{\text{max}}, \forall i\}$, where $d_{\text{max}}$ is a variable to be optimized. We can find that in such a subset the elements $d_i$, $\forall i$, are decoupled and constrained separately by $0\leq d_i\leq d_{\text{max}}$, for which we can simply use a scaled $\operatorname{Sigmoid}$ as the activation function. It should be pointed out that this approach may affect the optimality of the learned $\mathbf{d}$ depending on whether the optimal $\mathbf{d}$ is out of the subset or not. We will revisit this issue later in this subsection.

Now let us explore how to optimize the variable $d_{\text{max}}$ for the subset. Intuitively, it is desirable to maximize the region of the subset so as to increase the probability of containing the optimal $\mathbf{d}$. This can be translated into the optimization problem that maximizes  $d_{\text{max}}$ so that \eqref{E:central_point-d} holds for all $d_i$ satisfying $0\leq d_i\leq d_{\text{max}}$, $\forall i$, i.e.,
\begin{subequations} \label{E:problem-find}
    \begin{align}
    \max\  & d_{\text{max}} \label{E:problem1-Cd1} \\
    s.t.\  &  \mathbf{p}^C(\mathbf{d})\preceq P_{\text{max}}\mathbf{1},\ \forall \mathbf{d}\in\{\mathbf{d}|\mathbf{0}\preceq \mathbf{d}\preceq d_{\text{max}}\mathbf{1}\}. \label{E:problem-Cd3}
    \end{align}
\end{subequations}

To solve problem \eqref{E:problem-find}, we first prove in Appendix \ref{a:1} a property of the function  $\mathbf{p}^C(\mathbf{d})$ as follows.

\textit{\textbf{Proposition 1:} $\mathbf{p}^C$ satisfies $\mathbf{p}^C(\mathbf{d}')\preceq\mathbf{p}^C(\mathbf{d})$ if $\mathbf{d}'\preceq\mathbf{d}$.}

With Proposition 1, we know from \eqref{E:problem-Cd3} that $\mathbf{p}^C(\mathbf{d})\preceq \mathbf{p}^C(d_{\text{max}}\mathbf{1})$ holds, then constraint \eqref{E:problem-Cd3} can be simplified as
\begin{align} \label{E:cons-bound}
  \mathbf{p}^C(d_{\text{max}}\mathbf{1}) = \mathbf{p}_{0}+d_{\max}\mathbf{B}^{-1} \operatorname{diag}\left(\mathbf{B B}^{T}\right)^{\frac{1}{2}} \mathbf{1} \preceq P_{\text{max}}\mathbf{1},
\end{align}
where the equality follows from \eqref{E:central_point}. By replacing \eqref{E:problem-Cd3} with \eqref{E:cons-bound}, the maximal value of $d_{\text{max}}$ can be readily solved from problem~\eqref{E:problem-find}~as
\begin{equation}\label{E:dmax}
d_{\max}^\star=\min \left\{\frac{\left[P_{\text{max}}\mathbf{1}-\mathbf{p}_{0}\right]_i}{\left[\mathbf{B}^{-1} \operatorname{diag}(\mathbf{B B}^{T})^{\frac{1}{2}}\mathbf{1}\right]_i}, i=1,\dots,K\right\},
\end{equation}
where $[\mathbf{x}]_i$ denotes the $i$-th element of vector $\mathbf{x}$.
Note that $d_{\max}^\star$ is a constant that can be readily computed with \eqref{E:dmax} for any given channels and SINR requirements.

\begin{figure}
	\centering
	\includegraphics[width=0.75\textwidth]{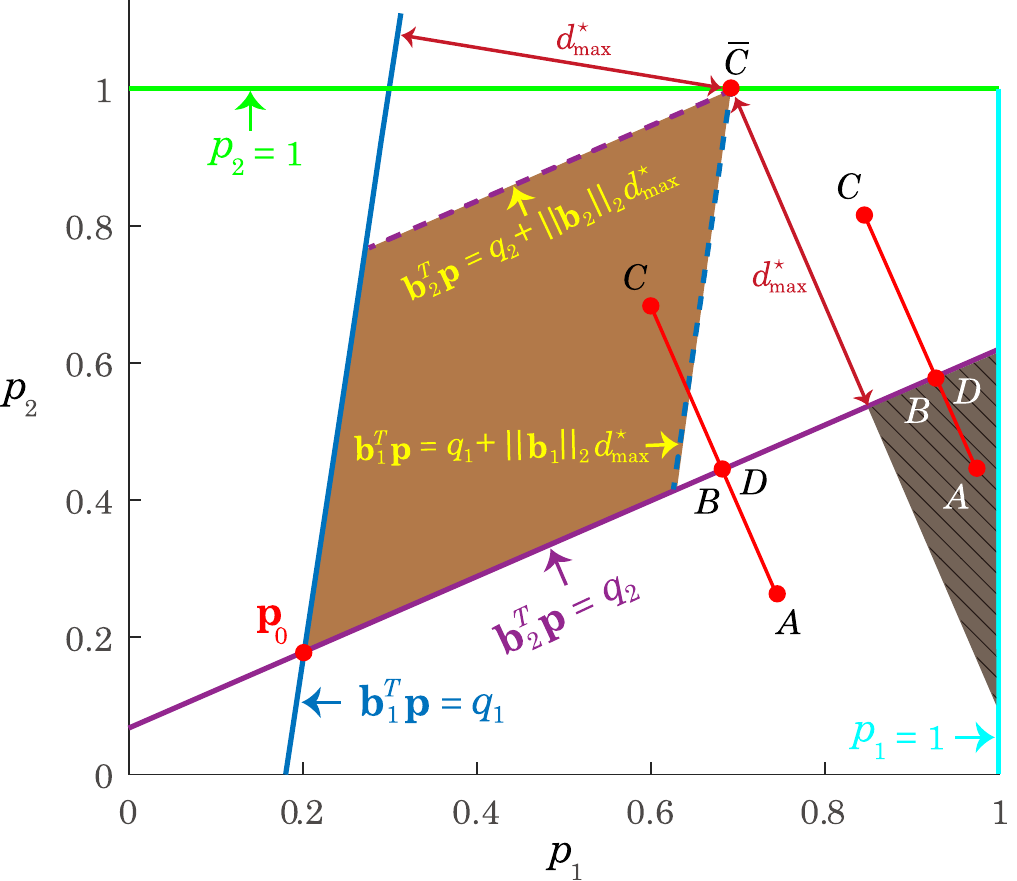}
    \caption{Illustration of the approach to find point $C$.}
	\label{fig:Cregion}
\end{figure}

In Fig.~\ref{fig:Cregion}, we illustrate the impact of the considered subset of $\mathbf{d}$ on point $C$, where the same setup as Fig.~\ref{fig:projection} is considered. Recall that
the position of point $C$ is determined by the distances $d_1$ and $d_2$. By substituting \eqref{E:dmax} into \eqref{E:cons-bound}, it is easy to find that the point with the distances $d_1=d_2=d_{\text{max}}^\star$, denoted by point $\bar{C}$, has at least one element equaling to $P_{\text{max}}$, which means that point $\bar{C}$ is located at the boundary of the feasible region of $\mathbf{p}$, as shown in Fig.~\ref{fig:Cregion}. From point $\bar{C}$, we can draw the two hyperplanes in parallel with the hyperplanes $\mathbf{b}_1^T\mathbf{p}=q_1$ and $\mathbf{b}_2^T\mathbf{p}=q_2$. The intersection of the four hyperplanes (shown shaded) is the region for finding point $C$, within which the distances $\mathbf{d}$ for every point satisfy $\{0\leq d_i\leq d_{\text{max}}^\star, \forall i\}$.

Compared with Fig.~\ref{fig:projection} where point $C$ is searched from the whole feasible region, the searching region of point $C$ shown in Fig.~\ref{fig:Cregion} shrinks. As aforementioned, the optimality of considering the shrunken searching space of point $C$ depends on whether the optimal point $C$ is included in the shaded area or not. Instead of rigorously analyzing the location of the optimal point $C$, which is very difficult, we attempt to give an intuitive geometrical description of the optimal location.
Taking Fig.~\ref{fig:Cregion} as an example, by using $\operatorname{Sigmoid}$ as the activation function for the output nodes corresponding to $\hat{\mathbf{p}}$, we can restrict $\hat{\mathbf{p}}$, denoted by point $A$, in the square ${0\leq p_1, p_2\leq1}$. If $\hat{\mathbf{p}}$ is an infeasible solution, then point $A$ is located outside the feasible region. The infeasible region can be divided into two parts: the line-filling shaded area and the remainder. If point $A$ is not located in the line-filling shaded area, then we can always find an optimal point $C$ in the shrunken searching space, which makes point $B$ and $D$ identical. If point $A$ is in the line-filling shaded area, then we can find that the optimal point $C$ is out of the shrunken searching space. Considering that the line-filling shaded area is much smaller than the whole infeasible area, it is of high probability that the shrunken searching space contains the optimal point $C$.

In summary, we learn the distance $\mathbf{d}$ through the second-half nodes of the output layer as shown in Fig.~\ref{fig:network}. We use the scaled $\operatorname{Sigmoid}$ function by $d_{\text{max}}^\star$ as the activation function for these nodes, which ensures that the output of each node satisfies $0\leq d_i\leq d_{\text{max}}^\star$, $\forall i$, where $d_{\text{max}}^\star$ can be computed from \eqref{E:dmax}. Meanwhile, for the first-half nodes of the output layer corresponding to $\hat{\mathbf{p}}$, the activation function is set as the scaled $\operatorname{Sigmoid}$ function by $P_{\text{max}}$, which restricts the infeasible region where point $A$ is located.
Upon substituting the learned distances $\mathbf{d}$ into \eqref{E:central_point}, the power of the interior point $C$ can be obtained.

\subsection{Finding Point $D$ and $E$}
Point $D$ is the cross point between the segment $AC$ and the boundary of the feasible region defined by constraint \eqref{E:proj-B2} and \eqref{E:proj-B3}. The segment $AC$ can be expressed as
\begin{align} \label{E:pointD}
    \mathbf{p} = \hat{\mathbf{p}} + \epsilon(\mathbf{p}^C-\hat{\mathbf{p}}),\ 0\leq\epsilon\leq1.
\end{align}
To find the cross point $D$ is equivalent to find the minimal $\epsilon$ that makes $\mathbf{p}$ given by \eqref{E:pointD} satisfy constraint \eqref{E:proj-B2} and \eqref{E:proj-B3}.

Since constraint \eqref{E:proj-B2} holds for $\mathbf{p}^C$ but not for $\hat{\mathbf{p}}$, there must exist a solution of $\epsilon$ with $0\leq\epsilon\leq1$ to make
the power $\mathbf{p}$ given by \eqref{E:pointD} satisfy \eqref{E:proj-B2}. It means that we do not need to explicitly consider the constraint $0\leq\epsilon\leq1$. Furthermore, since both $\mathbf{p}$ and $\mathbf{p}^C$ satisfy constraint \eqref{E:proj-B3} and $0\leq\epsilon\leq1$, we know that $\mathbf{p}$ given by \eqref{E:pointD} must satisfy \eqref{E:proj-B3}. Thus, it is also not necessary to consider constraint \eqref{E:proj-B3}. As a result, we can find the minimal $\epsilon$ by solving the following~problem
\begin{subequations} \label{E:proj-eps}
  \begin{align}
    \min_{\epsilon}\  & \epsilon \label{E:proj-problem1-1-eps} \\
    s.t.\  &  \mathbf{B}\left(\hat{\mathbf{p}} + \epsilon(\mathbf{p}^C-\hat{\mathbf{p}})\right) \succeq \mathbf{q}, \label{E:proj-B2-eps}
  \end{align}
\end{subequations}
where constraint \eqref{E:proj-B2-eps} comes from \eqref{E:proj-B2} by replacing $\mathbf{p}$ with \eqref{E:pointD}.

To minimize $\epsilon$, we are only interested in the lower bounds for $\epsilon$ constrained by \eqref{E:proj-B2-eps}. Then, by using $\mathcal{I}$ to denote the set of the indices of the positive elements in $\mathbf{B}(\mathbf{p}^C-\hat{\mathbf{p}})$, 
we can obtain from \eqref{E:proj-B2-eps} 
that
\begin{align}
\epsilon\geq \frac{[\mathbf{q}-\mathbf{B}\hat{\mathbf{p}}]_i}{[\mathbf{B}(\mathbf{p}^C-\hat{\mathbf{p}})]_i}, \forall i\in\mathcal{I},
\end{align}
from which the optimal $\epsilon^\star$ can be obtained as
\begin{align} \label{E:eps}
\epsilon^\star = \max\left\{\frac{[\mathbf{q}-\mathbf{B}\hat{\mathbf{p}}]_i}{[\mathbf{B}(\mathbf{p}^C\!-\!\hat{\mathbf{p}})]_i}, i\!\in\!\mathcal{I} 
\right\}.
\end{align}

By substituting $\epsilon^\star$ into \eqref{E:pointD}, we can obtain the power corresponding to point $D$, denoted by $\mathbf{p}^D$. By scaling $\mathbf{p}^D$, we can obtain the power of point $E$ as $\mathbf{p}^E=\frac{P_{\text{max}}}{\max\{\mathbf{p}^D\}}\mathbf{p}^D$.


In Table~\ref{algo}, we summarize the detailed procedures of the proposed projection block. The method only requires basic matrix operations without iterations and thus is of low complexity.
\begin{table}
\caption{Procedures of the Proposed Projection Block}
\label{algo}
	\hrule
    \begin{algorithmic}[1]
    \REQUIRE $\hat{\mathbf{p}}$ and $\mathbf{d}$ from the output layer.
    \ENSURE $\mathbf{p}$ satisfying per-user rate constraints and per-BS power constraints.
    \STATE 	Compute $d_{\max}^\star$ with \eqref{E:dmax}, and set $d_{\max}^\star\operatorname{Sigmoid}(\cdot)$ as the activation function for $\mathbf{d}$.
\STATE Find point $C$: substituting $\mathbf{d}$ into \eqref{E:central_point}, obtain the power corresponding to point $C$, i.e., $\mathbf{p}^C$.
    \STATE Find point $D$: compute $\epsilon^\star$ with \eqref{E:eps}, and obtain the power corresponding to point $D$ as $\mathbf{p}^D = \hat{\mathbf{p}} + \epsilon^\star(\mathbf{p}^C-\hat{\mathbf{p}})$.
    \STATE Find point $E$: scale $\mathbf{p}^D$ to obtain  $\mathbf{p}^E=\frac{P_{\text{max}}}{\max\{\mathbf{p}^D\}}\mathbf{p}^D$, and let the output $\mathbf{p}=\mathbf{p}^E$.
	\end{algorithmic}
    \hrule
\end{table}

\subsection{Gradient Backpropagation} \label{S:gradient}
In this subsection we derive the backpropagated gradient of the loss function $J_{\operatorname{SRNet}}$ with respect to the trainable parameters $\boldsymbol{\theta}$
of the DNN, including weights and bias, where the projection block is taken into~account.

With the loss function defined by \eqref{E:loss}, 
the parameters $\boldsymbol{\theta}$ can be updated for every epoch as
\begin{align} 
\boldsymbol{\theta} & \leftarrow \boldsymbol{\theta} -\frac{\eta}{M_{t r}} \sum_{m=1}^{M_{t r}} \frac{\partial J^{(m)}}{\partial\boldsymbol{\theta}},\label{E:wgtbias-B1} 
\end{align}
where $J^{(m)}$ is defined in \eqref{E:loss}, $M_{tr}$ is the batch size of an epoch for training, and $\eta$ is the learning rate. 

\begin{figure}
	\centering
	\includegraphics[width=0.8\textwidth]{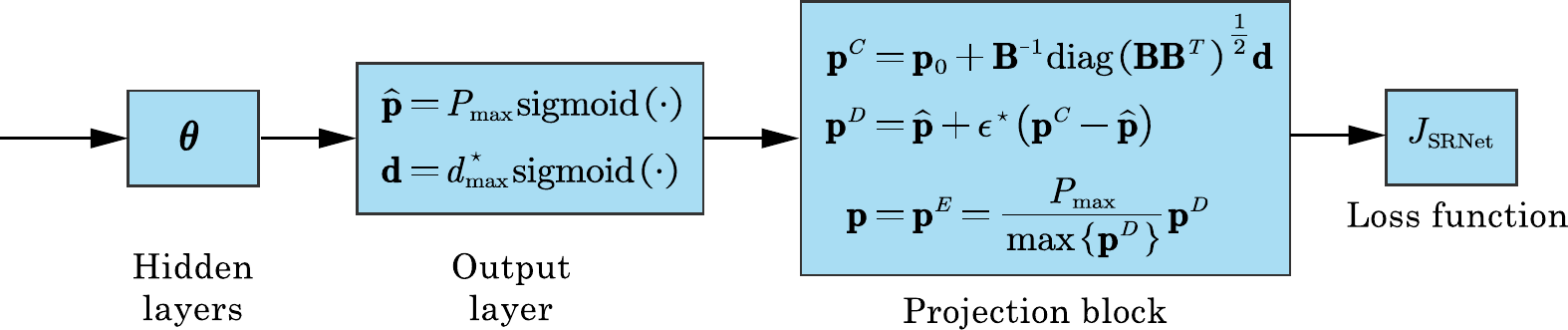}
    \caption{Illustration of the dependency among the related variables.}
	\label{fig:dependency}
\end{figure}

Next, we derive the gradient $\frac{\partial J}{\partial\boldsymbol{\theta}}$, where the index of samples $^{(m)}$ is omitted for notational simplicity. In Fig.~\ref{fig:dependency}, the dependency among the variables connecting $\boldsymbol{\theta}$ and $J$ is illustrated. Based on the chain rule of matrix derivative, $\frac{\partial J}{\partial\boldsymbol{\theta}}$ can be expanded as
\begin{equation}\label{E:weight}
\frac{\partial J}{\partial\boldsymbol{\theta}}=\frac{\partial J}{\partial \mathbf{p}^{E}} \frac{\partial \mathbf{p}^{E}}{\partial \mathbf{p}^{D}}\left(\frac{\partial \mathbf{p}^{D}}{\partial \hat{\mathbf{p}}} \frac{\partial \hat{\mathbf{p}}}{\partial \boldsymbol{\theta}}+\frac{\partial \mathbf{p}^{D}}{\partial \mathbf{p}^{C}} \frac{\partial {\mathbf{p}}^{C}}{\partial \mathbf{d}}\frac{\partial {\mathbf{d}}}{\partial \boldsymbol{\theta}}\right).
\end{equation}

Both $\hat{\mathbf{p}}$ and $\mathbf{d}$ are scaled $\operatorname{Sigmoid}$  function of $\boldsymbol{\theta}$, where the scale factor for $\hat{\mathbf{p}}$ is the constant $P_{\text{max}}$ and that for  $\mathbf{d}$ is $d_{\text{max}}^\star$ that can be computed with \eqref{E:dmax} for any given channels and rate requirements. Thus, the terms $\frac{\partial \hat{\mathbf{p}}}{\partial \boldsymbol{\theta}}$ and $\frac{\partial {\mathbf{d}}}{\partial \boldsymbol{\theta}}$ can be easily derived as for a traditional DNN. We next focus on the projection block and derive the remaining five terms, respectively. 

\subsubsection{$\frac{\partial J}{\partial \mathbf{p}^{E}}$}
Defining $g_{ij}=\alpha_{ij}\left|h_{ij}\right|^{2}$, we can obtain from \eqref{E:SINR} and \eqref{E:loss} that
\begin{equation}\label{E:BackRE}
\frac{\partial J}{\partial \mathbf{p}^{E}}=\frac{1}{\ln 2\left(\sum_{j=1}^{K} g_{ij} [\mathbf{p}^{E}]_{j}+\sigma^{2}\right)} \mathbf{c},
\end{equation}
where the $k$-th element of vector  $\mathbf{c}$ is
\begin{equation}
[\mathbf{c}]_{k}=g_{kk}+\sum_{i=1, i \neq k}^{K} \frac{-g_{ik} g_{i i}\left[\mathbf{p}^{E}\right]_{i}}{\sum_{j=1, j \neq i}^{K} g_{ij}\left[\mathbf{p}^{E}\right]_{j}+\sigma^{2}}.
\end{equation}

\subsubsection{$\frac{\partial \mathbf{p}^{E} }{ \partial \mathbf{p}^{D}}$}
Without loss of generality, assuming that the $k$-th element of $\mathbf{p}^{D}$ in the current gradient iteration is the maximum, we can obtain that $\mathbf{p}^E = \frac{P_{\text{max}}}{[\mathbf{p}^D]_{k}}\mathbf{p}^D$. With some regular manipulations, we can derive $\frac{\partial \mathbf{p}^{E} }{ \partial \mathbf{p}^{D}}$ as
 \begin{equation}\label{E:BackED}
 \frac{\partial \mathbf{p}^{E}}{\partial \mathbf{p}^{D}}
 = \frac{P_{\max}}{\left([\mathbf{p}^{D}]_{k}\right)^{2}}\mathbf{M},
 \end{equation}
 where matrix $\mathbf{M}$ is defined as
\begin{equation}\label{E:BackED-M}
[\mathbf{M}]_{i j}=\left\{\begin{array}{cc}
\left[\mathbf{p}^{D}\right]_{k},& i=j \text { and } i \neq k,\\
-{[\mathbf{p}^{D}]_{j}},& i \neq j \text { and } i=k, \\
0,&\text{else}.
\end{array}\right.
\end{equation}

\subsubsection{$\frac{\partial \mathbf{p}^{D}}{ \partial  \hat{\mathbf{p}}}$}
To derive this term, without loss of generality, assume that $\epsilon^\star$ equals to the $k$-th element in the right-hand side of \eqref{E:eps}, i.e.,
\begin{equation}\label{E:epsilon-max}
\epsilon^{\star}=\frac{[\mathbf{q}-\mathbf{B} \hat{\mathbf{p}}]_{k}}{\left[\mathbf{B}\left(\mathbf{p}^{C}-\hat{\mathbf{p}}\right)\right]_{k}}.
\end{equation}
Then, with the the derivation in Appendix~\ref{S:derivation}, we can obtain $\frac{\partial \mathbf{p}^{D}}{ \partial  \hat{\mathbf{p}}}$ as
\begin{equation}\label{E:BackDP}
\frac{\partial \mathbf{p}^{D}}{\partial \hat{\mathbf{p}}}=\frac{[\mathbf{Bp}^{C}-\mathbf{q}]_{k}}{\left[\mathbf{B}\left(\mathbf{p}^{C}-\hat{\mathbf{p}}\right)\right]_{k}}\mathbf{I}+\frac{\left[\mathbf{B}(\mathbf{p}^{C}-\mathbf{q})\right]_{k}}
{{\|[\mathbf{B}(\mathbf{p}^{C}-\hat{\mathbf{p}})]_{k}\|^{2}}} \widehat{\mathbf{M}},
\end{equation}
where  $\mathbf{I}$ is the identity matrix and $\widehat{\mathbf{M}}$ is defined as $[\widehat{\mathbf{M}}]_{ij} =[\hat{\mathbf{p}}-\mathbf{p}^{C}]_{j} [\mathbf{B}]_{ki}$.

\subsubsection{$\frac{\partial \mathbf{p}^{D}}{ \partial  \mathbf{p}_{C}}$}
As derived in Appendix~\ref{S:derivation}, $\frac{\partial \mathbf{p}^{D}}{ \partial  \mathbf{p}_{C}}$ can be obtained as
  \begin{equation}\label{E:BackDC}
   \frac{\partial \mathbf{p}^{D}}{\partial \mathbf{p}^{C}}=\frac{[\mathbf{q}-\mathbf{B} \hat{\mathbf{p}}]_{k}}{\left[\mathbf{B}\left(\mathbf{p}^{C}-\hat{\mathbf{p}}\right)\right]_{k}}\mathbf{I}
  -\frac{[\mathbf{q}-\mathbf{B} \hat{\mathbf{p}}]_{k}}
   {{\|[\mathbf{B}(\mathbf{p}^{C}-\hat{\mathbf{p}})]_{k}\|^{2}}}\overline{\mathbf{M}},
   \end{equation}
   where matrix $\overline{\mathbf{M}}$ is defined as $[\overline{\mathbf{M}}]_{ij} =[\mathbf{p}^{C}-\hat{\mathbf{p}}]_{j} [\mathbf{B}]_{ki}$.

\subsubsection{$\frac{\partial \mathbf{p}^{C}}{ \partial  \mathbf{d}}$}
With \eqref{E:central_point}, $\frac{\partial \mathbf{p}^{C}}{ \partial  \mathbf{d}}$ can be easily obtained as
\begin{equation}\label{E:BackCd}
   \frac{\partial \mathbf{p}^{C}}{\partial \mathbf{d}}= \operatorname{diag}\left(\mathbf{B B}^{T}\right)^{\frac{1}{2}}(\mathbf{B}^{T})^{-1}.
\end{equation}

Upon substituting these five partial derivatives into \eqref{E:weight}, the backpropagated gradient can be obtained, and then we can directly maximize the sum rate by training the parameters of the DNN in an unsupervised learning manner.

\subsection{A Heuristic Selection of Point $C$}\label{section:4}

In the previous subsections, we have described the full design of the employed DNN and the projection block, where the critical point $C$ is obtained via learning the distance $\mathbf{d}$. In this subsection we provide a heuristic selection of point $C$ without resorting to learning, which is able to reduce the dimension of the output layer of the DNN. 

Let us describe the basic idea with Fig.~\ref{fig:projection}. Comparing the two projection points $D$ and $D'$, which correspond to the interior points $C$ and $C'$, respectively, we can find that point $D$ is closer to the $\ell_2$ projection point $B$. It indicates that point $C$ is better than point $C'$ for this instance. Now let us consider that point $A$ is randomly located in the infeasible area, i.e., the white area within the square defined by $0\leq p_1, p_2\leq 1$.
It is not difficult to find that setting point $C$ away from the boundary of the feasible region is robust in the sense of avoiding a large projection error between point $B$ and $D$. In the example shown in Fig.~\ref{fig:projection}, point $C'$ is located close to the bottom-left boundary, which will lead to a large projection error for point $A$ located at the top-left or bottom-right corner. The observation motivates us to set point $C$ as the one with the largest distance from the boundaries of the feasible region. Since we use the $\operatorname{Sigmoid}$ function to restrict point $A$ to be located in the area $\mathbf{0} \preceq \mathbf{p} \preceq P_{\text{max}}\mathbf{1}$, we only need to consider the boundaries defined by constraints $\mathbf{B}\mathbf{p} \succeq \mathbf{q}$. As a result, finding point $C$ is equivalent to maximizing the minimum of distance $\mathbf{d}$ between point $C$ and the boundary hyperplanes $\mathbf{B}\mathbf{p} = \mathbf{q}$. The optimization problem can be formulated~as
\begin{subequations} \label{E:problem-appro}
    \begin{align}
    \max_{\mathbf{d}}\  & \min \{\mathbf{d}\} \label{E:problem1-C1} \\
    s.t.\  &  \mathbf{d} \succeq \mathbf{0}, \label{E:problem-C2}\\
           &   \mathbf{p}^C(\mathbf{d})\preceq P_{\text{max}}\mathbf{1}, \label{E:problem-C3}
    \end{align}
\end{subequations}
where $\mathbf{p}^C(\mathbf{d})$ is the power corresponding to point $C$ given by \eqref{E:central_point}, and  constraint \eqref{E:problem-C3} ensures that $\mathbf{p}^C(\mathbf{d})$ is inside the feasible area.


We prove in Appendix~\ref{a:2} that the optimal solution to problem \eqref{E:problem-appro} is
\begin{align}
  \mathbf{d}=d_{\max}^\star\mathbf{1},
\end{align}
where $d_{\max}^\star$ is given by \eqref{E:dmax}. The corresponding power of point $C$ is
\begin{align} \label{E:cons-bound-heu}
  \mathbf{p}^C = \mathbf{p}_{0}+d_{\max}^\star\mathbf{B}^{-1} \operatorname{diag}\left(\mathbf{B B}^{T}\right)^{\frac{1}{2}} \mathbf{1}.
\end{align}

\begin{figure}
	\centering
	\includegraphics[width=0.8\textwidth]{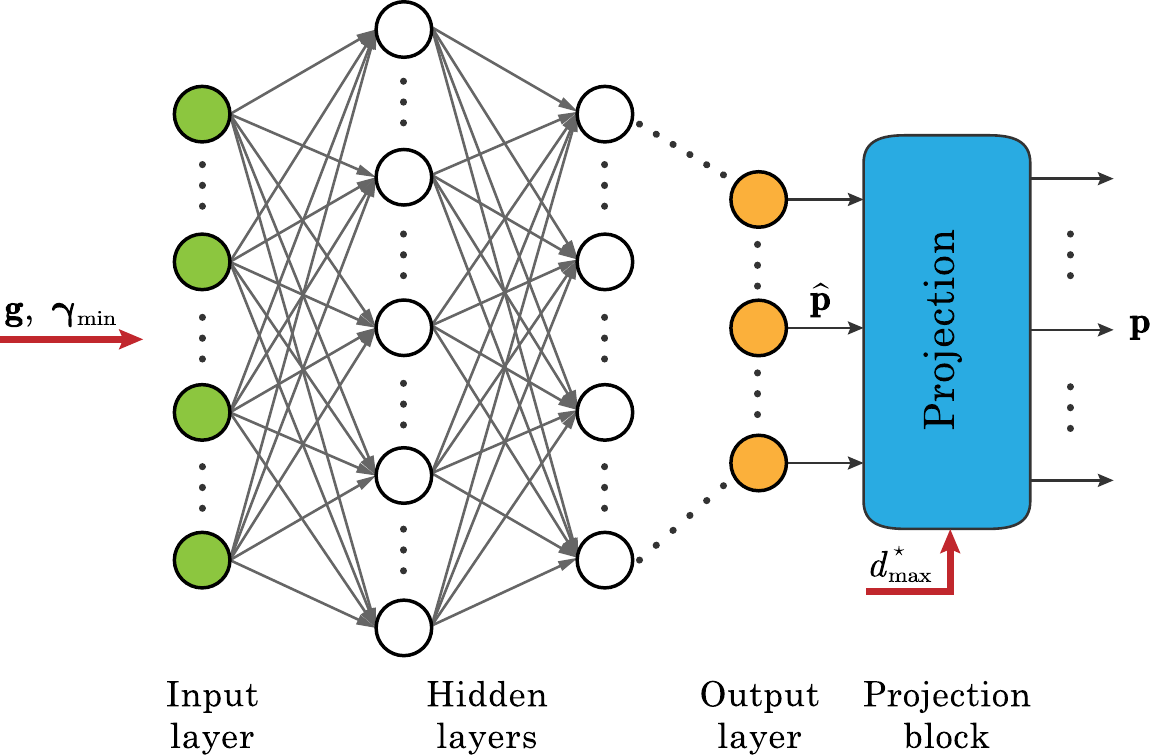}
     \caption{Structure of the employed DNN with a heuristic selection of point $C$.}
	\label{fig:network-heu}
\end{figure}

With the heuristic selection of point $C$, the structure of the employed DNN can be simplified as Fig.~\ref{fig:network-heu}, where the number of nodes of the output layer reduces to $K$, the number of inputs to the projection block reduces to $K+1$, and $d_{\max}^\star$ is introduced as an input to the projection block, which is used to compute $\mathbf{p}^C$ with \eqref{E:cons-bound-heu}. The backpropagated gradient is also simplified because now $\mathbf{d}=d_{\max}^\star\mathbf{1}$ is a constant, leading to $\dfrac{\partial \mathbf{d}}{\partial \boldsymbol{\theta}}=0$ and $\dfrac{\partial J}{\partial \boldsymbol{\theta}}$ as
\begin{equation}\frac{\partial J}{\partial \boldsymbol{\theta}}=\frac{\partial J}{\partial \mathbf{p}^{E}} \frac{\partial \mathbf{p}^{E}}{\partial \mathbf{p}^{D}}\frac{\partial \mathbf{p}^{D}}{\partial \hat{\mathbf{p}}} \frac{\partial \hat{\mathbf{p}}}{\partial \boldsymbol{\theta}},
\end{equation}
which can be computed with the results of Sec.~\ref{S:gradient}.

\section{Performance Evaluation}
\label{s:5}

In this section, we evaluate the performance of the proposed SRNet. Consider the downlink transmission of $K=3$ adjacent cells, as shown in Fig.~\ref{fig:network-eva}, where the cell radius is $250$~m and the maximal transmit power of BS is $P_{\text{max}}=46$~dBm. The noise power is $\sigma^2=-92$~dBm, which corresponds to the noise power spectral density of -174~dBm/Hz, bandwidth of 20 MHz, and noise figure of 9~dB~\cite{TR36.814}.

\subsection{Methods for Comparison}
The methods for comparison are described as follows, where  WMMSE is not considered because it cannot be applied for the case with per-user rate constraints.
\begin{itemize}
  \item[1)] \textbf{SRNet:} This is the proposed deep learning method with the following fine-tuned hyper-parameters and configurations.
      Consider four hidden layers for SRNet, where the number of nodes is $[720,360,180,90]$. For every hidden layer, we use $\operatorname{ReLU}$ as the activation function and add  the batch normalization layer~\cite{Sergey2015} before activation function. For the output layer, we use the scaled $\operatorname{Sigmoid}$ by $P_{\max}$ and $d_{\max}^\star$ as the activation functions for the nodes corresponding to $\hat{\mathbf{p}}$ and $\mathbf{d}$ in order to limit $\hat{\mathbf{p}}$ and $\mathbf{d}$ within $[0, P_{\max}]$ and $[0, d_{\max}^\star]$, respectively.
      We use the mini-batch gradient descent method to train the network~\cite{Goodfellow2016}, where the weights and biases are initialized by the Xavier method~\cite{Xavier2010}. The number of iterations is set as $1.5\!\times\!10^{5}$ and the batch size for every iteration is $8000$. After trying the gradient descent methods including AdaGrad, RMSprop and Adam~\cite{kingma2014adam}, we choose Adam to adaptively adjust the learning~rate.
  \item[2)] \textbf{SRNet-Heu:} This is SRNet with the proposed heuristic method for selecting point $C$, as shown by Fig.~\ref{fig:network-heu}.
  \item[3)] \textbf{PCNet, ePCNet:} PCNet is the method proposed by~\cite{Liang2018Towards}, which addresses the per-user rate constraints by adding the penalty of the constraint
violation to the loss function. In simulations, we have carefully searched the penalty parameter to achieve good performance. We have also adjusted the structure of the DNN used in PCNet, e.g., the numbers of layers and nodes, activation function, etc., and selected the best one based on test results. The output of the method may not satisfy the constraints. If this happens, as \cite{Liang2018Towards} suggests, the solution $\mathbf{p} = \mathbf{B}^{-1}\mathbf{q}$ is used as the final output. In \cite{Liang2018Towards}, ensemble learning is used to enhance the performance of PCNet, namely ePCNet, which trains a number of PCNet and selects the one providing the highest sum rate. We set the ensemble size as $10$.
\begin{figure}
	\centering
	\includegraphics[width=0.65\textwidth]{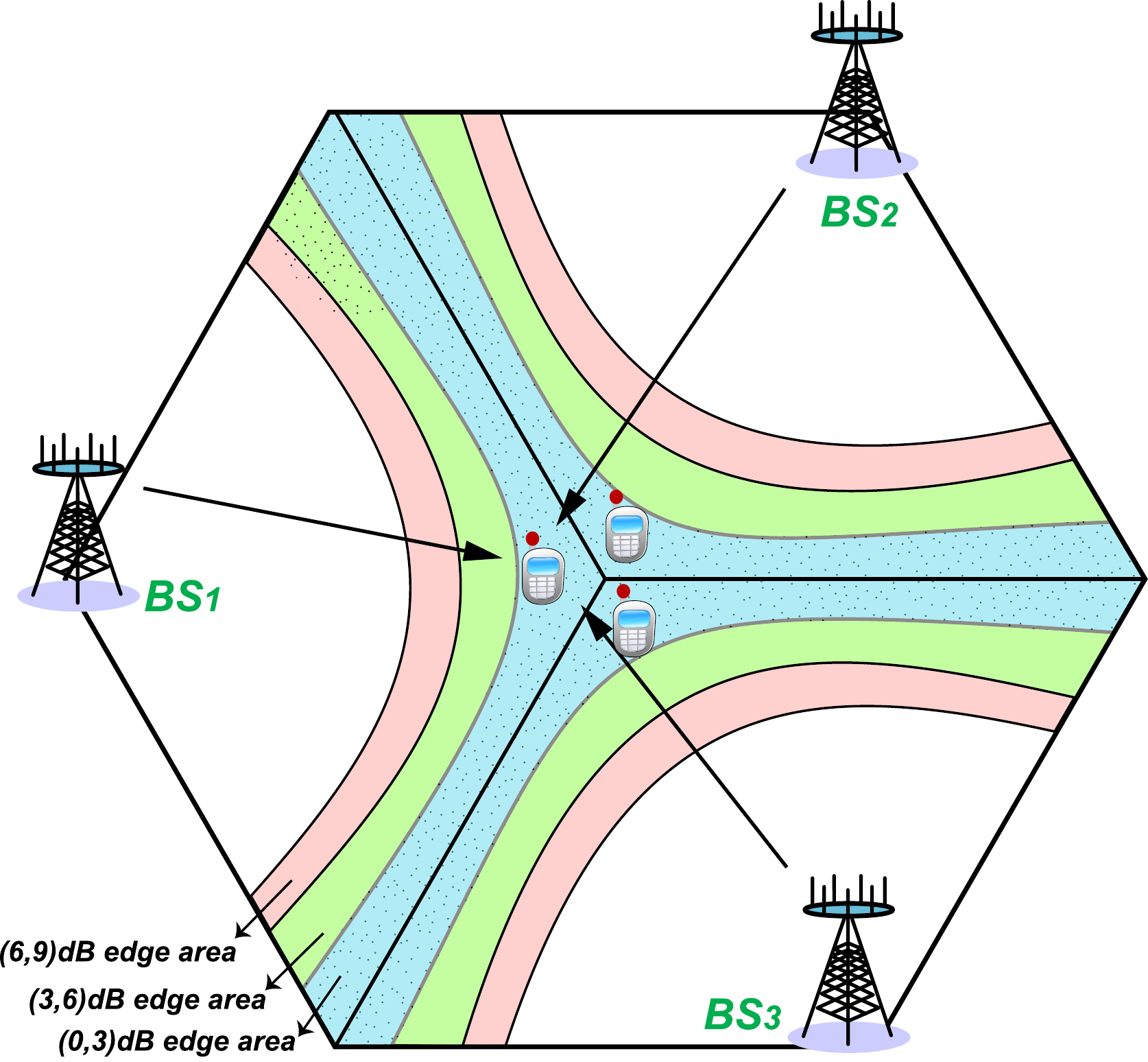}
     \caption{Network layout and cell-edge region for performance evaluation.}
	\label{fig:network-eva}
\end{figure}
\item[4)] \textbf{PC-M:} This is the method proposed by~\cite{Lee2020}, which addresses the constraints by multiplying the loss function with a penalty of constraint violation. We have also adjusted the parameters of the DNN used in this method and selected the best one. If the output of the method can not satisfy the constraints, the solution $\mathbf{p} = \mathbf{B}^{-1}\mathbf{q}$ is used as the final output as~PCNet.
  \item[5)] \textbf{SQP, Interior Point: } SQP and Interior Point are two widely used numerical optimization algorithms~\cite{Nocedal2006}, which can be used to solve the non-convex SRM problem. In simulations, we take use of $\operatorname{fmincon}$ function in Matlab to implement the two algorithms.
\end{itemize}

\subsection{Training and Testing Datasets}

We consider different training and testing datasets in simulations for different purposes. The inputs of SRNet include channel gains $\alpha_{ij}|h_{ij}|^2$ and rate requirements $r_{i,\text{min}}$ (or equivalently SINR requirements $\gamma_{i,\text{min}}$), with respect to which we generate the following training datasets.

\begin{itemize}
  \item[1)] \textbf{Train-$(\rho_{\min},\rho_{\max})$dB-$\lambda$:} In the dataset,
  the UEs are uniformly located in  a so-called ``$(\rho_{\min},$ $\rho_{\max})$dB cell-edge area'', which is defined for the $i$-th cell as the region where $\rho_{\min}\leq\alpha_{ii}-\max_{j\neq i}\{\alpha_{ij}\} < \rho_{\max}$~dB. Fig.~\ref{fig:network-eva} illustrates the $(0,3)$~dB, $(3,6)$~dB and $(6,9)$~dB cell-edge regions, where the region boundary is smooth because the shadowing is not considered here for better illustration while it is considered in simulations. The pathloss is modeled by $36.3+37.6\log d$ with $d$ denoting the distance between a BS and a UE, and the shadowing follows the log-normal distribution with the standard deviation of 8~dB~\cite{TR36.814}. The small-scale channels follow Rayleigh fading.
      The minimum rate requirements of UEs are set as the same, i.e., $r_{i,\text{min}} = \lambda, \forall i$. For any given $(\rho_{\min},\rho_{\max})$ and $\lambda$, we generate channel samples by randomly dropping UEs to compute large-scale fading gains $\alpha_{ij}$ and independently generating small-scale fading channel gains $|h_{ij}|^2$. For each generated channel sample, we check if the minimum rate requirement $\lambda$ is achievable, which can be done by the feasibility judgement method presented in Sec.~\ref{s:projection}. If it is unachievable, which means that the SRM problem is infeasible under the channel sample, then we discard this sample since it calls for admission control of UEs that is beyond the scope of the paper. Otherwise, it is saved as a feasible sample. We totally generate $4\!\times\!10^{6}$ feasible channel samples in the training dataset for each $(\rho_{\min},\rho_{\max})$ and $\lambda$.
      In simulations, we consider $(\rho_{\min},\rho_{\max})=(0,3)$ and $(6,9)$~dB, and $\lambda=0.1, 0.2, 0.3, 0.4$ and $0.5$~bit/s/Hz, respectively, resulting in 10 training datasets in total.
  \item[2)] \textbf{Train-$(\rho_{\min},\rho_{\max})$dB-Random:} The difference between this dataset and the previous datasets lies in the minimum rate requirements of UEs $r_{i,\text{min}}$, $\forall i$. Now $r_{i,\text{min}}$ is randomly sampled from $0.1$ to $1.0$~bit/s/Hz with the step of $0.1$~bit/s/Hz, in which each sample has an equal probability of being chosen. Given the selected $r_{i,\text{min}}$, $\forall i$, we generate a feasible channel sample, which needs to ensure the selected rate requirement to be achievable. By repeating the random generation of rate requirements and channels, we finally obtain $4\!\times\!10^{6}$ feasible training samples.
\end{itemize}

The used testing dataset is called \textbf{Test-$(\rho_{\min},\rho_{\max})$dB-$\lambda$}, which is independently generated by the same approach as the training dataset \textbf{Train-$(\rho_{\min},\rho_{\max})$dB-$\lambda$} and consists of $10^{4}$ samples for performance evaluation.

\subsection{Evaluation Results}
\label{s:5.2}

\subsubsection{\underline{Performance Comparison}}
\begin{figure}
	\centering  
	\includegraphics[width=0.75\textwidth]{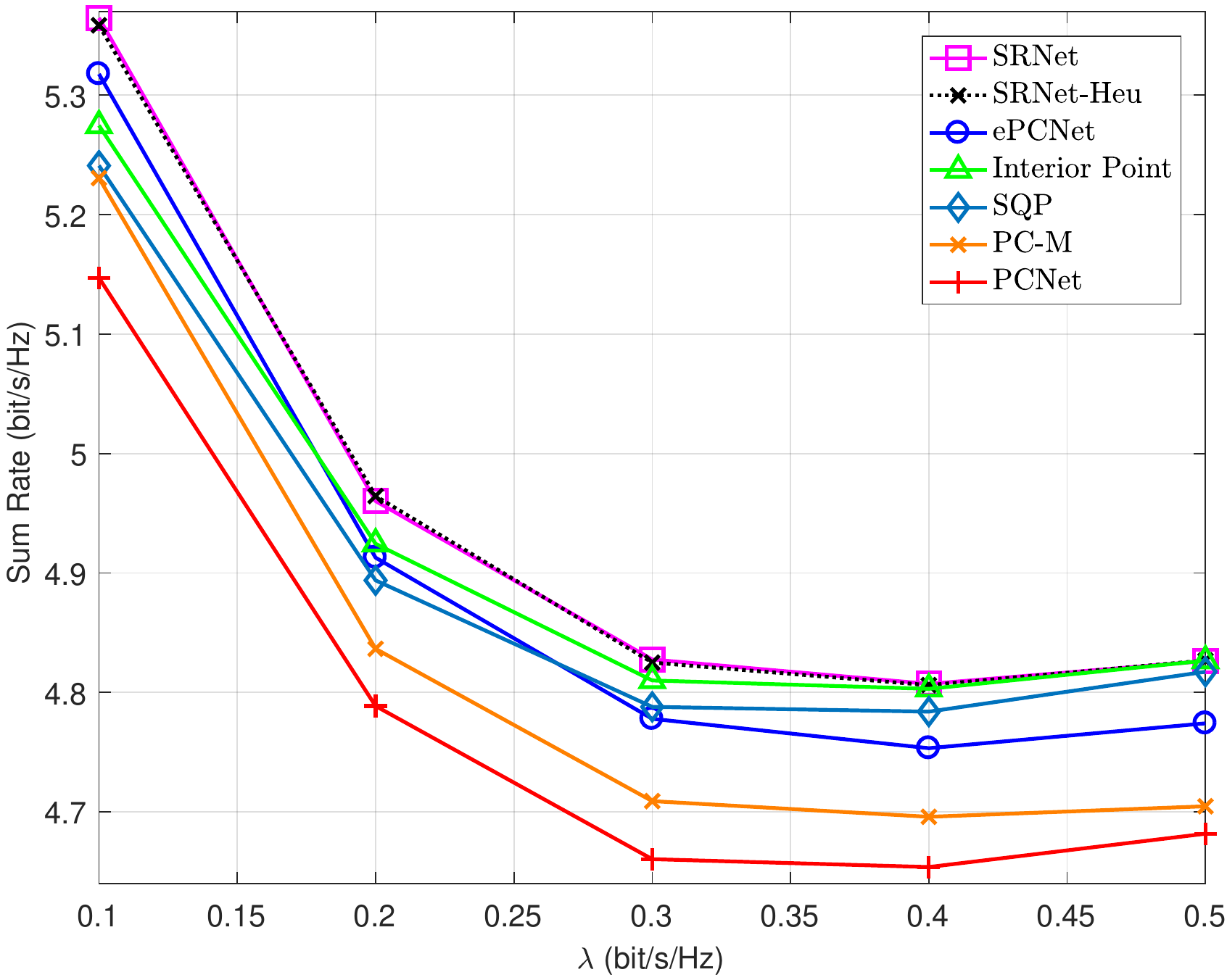}  
	\caption{Performance comparison for different rate requirements $\lambda$.}  
    \label{fig6}   
\end{figure}

We evaluate the performance of the compared methods by using the dataset \textbf{Train-$(0,3)$dB-$\lambda$} for training and \textbf{Test-$(0,3)$dB-$\lambda$} for testing, where the two datasets use the same rate requirement $\lambda$.
Fig.~\ref{fig6} shows the sum rate as a function of $\lambda$, where the UEs are located in $(0,3)$dB cell edge area. For every deep learning based method, we re-train the network whenever $\lambda$ changes. The proposed methods can always satisfy the per-user rate constraints, while PCNet, ePCNet and PC-M may not satisfy the constraints for some testing samples, for which the power control is computed as $\mathbf{p} = \mathbf{B}^{-1}\mathbf{q}$ as aforementioned.

We can find that the sum rate first decreases with $\lambda$ from 0.1 to 0.4, and then increases from 0.4 to 0.5. The result may appear counterintuitive since a higher $\lambda$ means a stricter rate constraint, which should lead to the decrease of sum rate. Nevertheless, it can be explained by considering another impact of $\lambda$ on the sum rate. Specifically, when $\lambda$ is too high, the rate constraints are only feasible for some good channels. Thus, the feasible channels for $\lambda\!=\!0.5$ are statistically better than those for $\lambda\!=\!0.4$, which results in an improved sum rate when increasing $\lambda$ from $0.4$ to $0.5$. It can be found from Fig.~\ref{fig6} that the proposed SRNet achieves the best performance, which outperforms existing deep learning methods PCNet and PC-M. By selecting the best one from 10 PCNets, ePCNet can achieve better performance as expected, which, nevertheless, is still inferior to SRNet. Both SQP and Interior Point are inferior to SRNet for low rate requirements, and the gap between them reduces with the increase of rate requirement as the feasible region shrinks when the constraints become stricter. It is also shown that SRNet-Heu performs close to SRNet, indicating that the proposed heuristic selection of point $C$ works well.

\begin{figure*}
	\centering
    \subfigure[Constraint satisfaction comparison]{
        \includegraphics[width=0.47\textwidth]{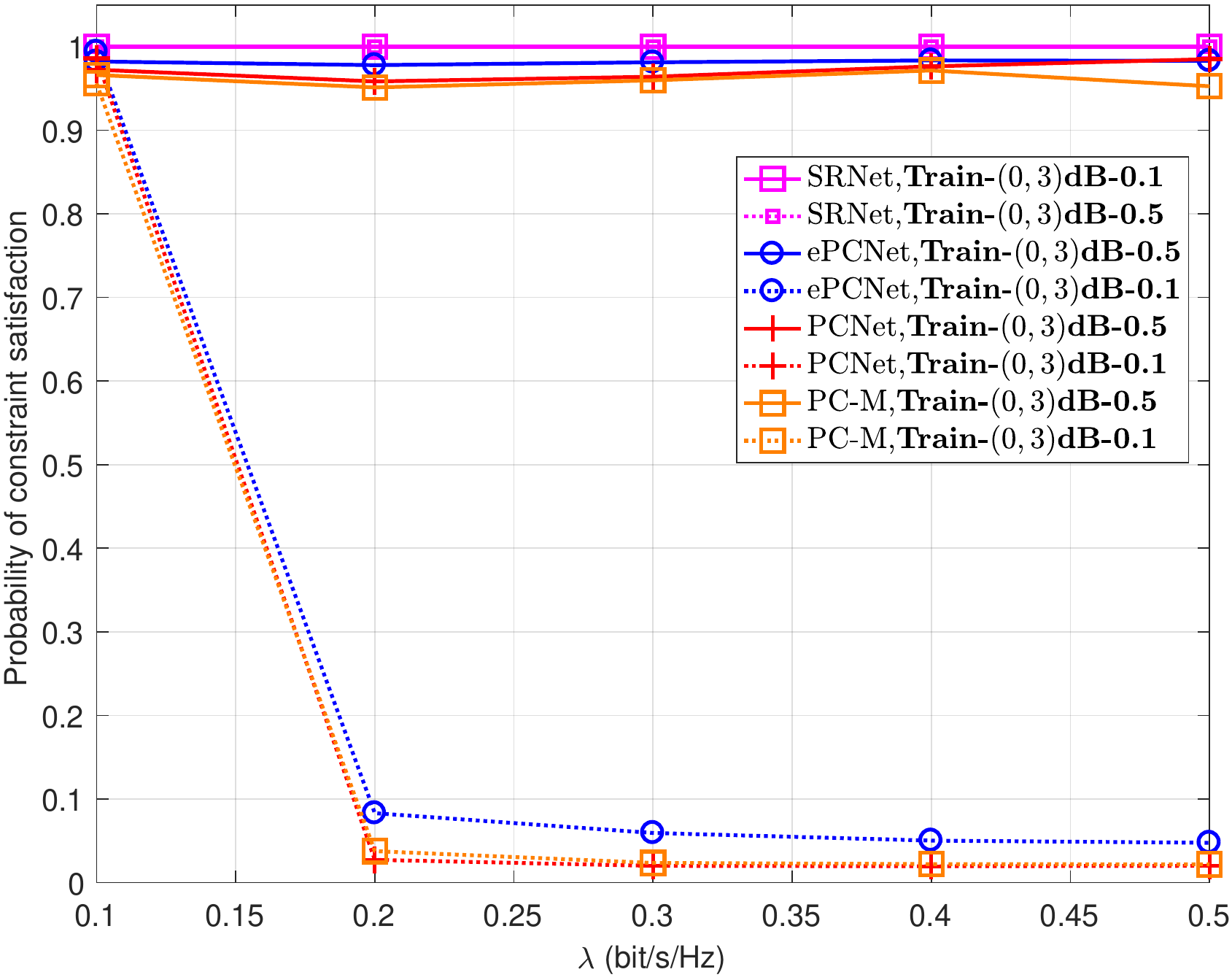} \label{F:constraint}
        }
    \subfigure[Performance comparison]{
        \includegraphics[width=0.47\textwidth]{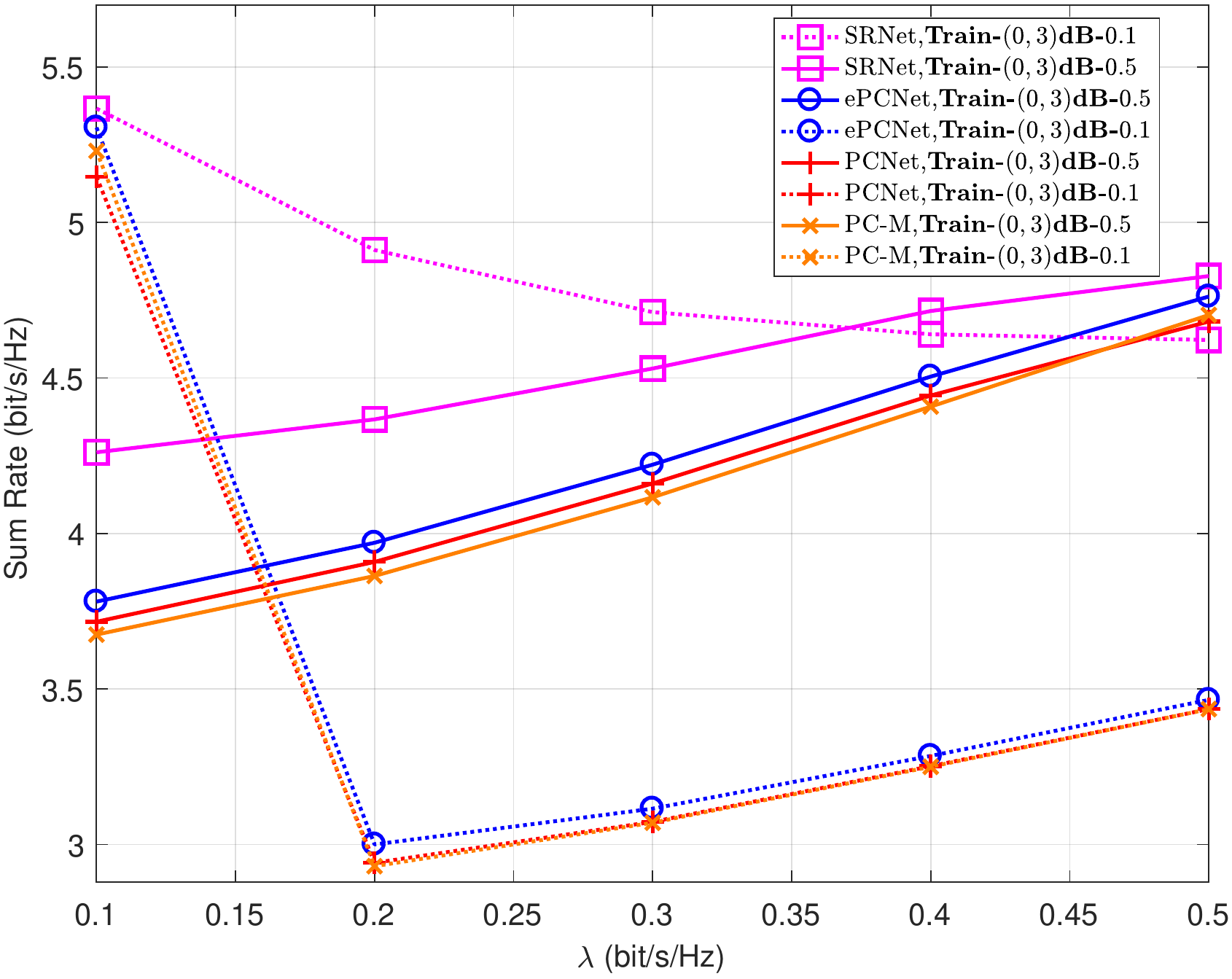} \label{F:sumrate}
        }
	\caption{Constraint satisfaction and performance comparison under rate requirement mismatch.}\label{F:mismatch}
\end{figure*}


We next compare the satisfaction of the per-user rate constraints and performance of the deep learning methods, where the impact of mismatch of training and testing datasets is taken into account. In Fig.~\ref{F:mismatch}, we train SRNet, PCNet, and PC-M with the datasets \textbf{Train-$(0,3)$dB-$\lambda'$} with $\lambda'=0.1$ or $0.5$, while conduct the testing with \textbf{Test-$(0,3)$dB-$\lambda$}. The training and testing datasets match if $\lambda' = \lambda$ while mismatch otherwise. Fig.~\ref{F:constraint} and \ref{F:sumrate} depict the probability of constraint satisfaction and the corresponding sum rate, respectively, where the rate requirements $\lambda$ for testing are placed in the $x$-axis and the rate requirements $\lambda'$ for training are shown in the~legend.

Let us first consider the proposed SRNet. SRNet ensures the constraints by cascading a projection block after the DNN, which does not contain trainable parameters. Thus, the mismatch between training and testing datasets does not affect the satisfaction of constraints.
It is shown by Fig.~\ref{F:constraint} that when applying the SRNet trained by \textbf{Train-$(0,3)$dB-$0.1$} or by \textbf{Train-$(0,3)$dB-$0.5$} to the testing dataset \textbf{Test-$(0,3)$dB-$\lambda$}, the constraints can be always satisfied for different $\lambda$. By comparing Fig.~\ref{F:sumrate} and Fig.~\ref{fig6}, we can observe the performance loss caused by model mismatch. As expected, the SRNets trained by \textbf{Train-$(0,3)$dB-$0.1$} and \textbf{Train-$(0,3)$dB-$0.5$} perform well for low and high $\lambda$, respectively, where the gap between them is large for low $\lambda$ and small for high $\lambda$. This is because \textbf{Train-$(0,3)$dB-$0.5$} only contains good channels while \textbf{Train-$(0,3)$dB-$0.1$} contains both good and bad channels, which makes the former suffer from a severer model mismatch at low $\lambda$ than the latter at high $\lambda$.

For PCNet, ePCNet, and PC-M, let us first focus on the case trained by \textbf{Train-$(0,3)$dB-$0.1$}. It is shown from Fig.~\ref{F:constraint} that when $\lambda = 0.1$, i.e., without model mismatch, by carefully selecting the penalty parameter for constraint violation, the constraints can be satisfied with a high probability (but still not fully satisfied). When $\lambda > 0.1$, i.e., with model mismatch, however, the constraints can be rarely satisfied. This leads to a quick drop of performance as shown by Fig.~\ref{F:sumrate}, where the sum rate increases when $\lambda$ grows from 0.2 to 0.5 because the feasible channels become better for higher $\lambda$ as analyzed in Fig.~\ref{fig6}. The results indicate that PCNet, ePCNet and PC-M trained for a given rate requirement is not applicable to the case with a higher rate requirement. Now let us examine the training under a high rate requirement, i.e., with \textbf{Train-$(0,3)$dB-$0.5$}. We can observe two-fold impact of doing so. On one hand, training PCNet, ePCNet and PC-M under a higher rate requirement can guarantee the lower rate requirements with a high probability, as shown by Fig.~\ref{F:constraint}. On the other hand, nevertheless, training the networks under strict constraints leads to conservative performance, and an evident performance gap between the three methods and SRNet can be observed from Fig.~\ref{F:sumrate}.

\begin{figure}
	\centering  
	\includegraphics[width=0.75\textwidth]{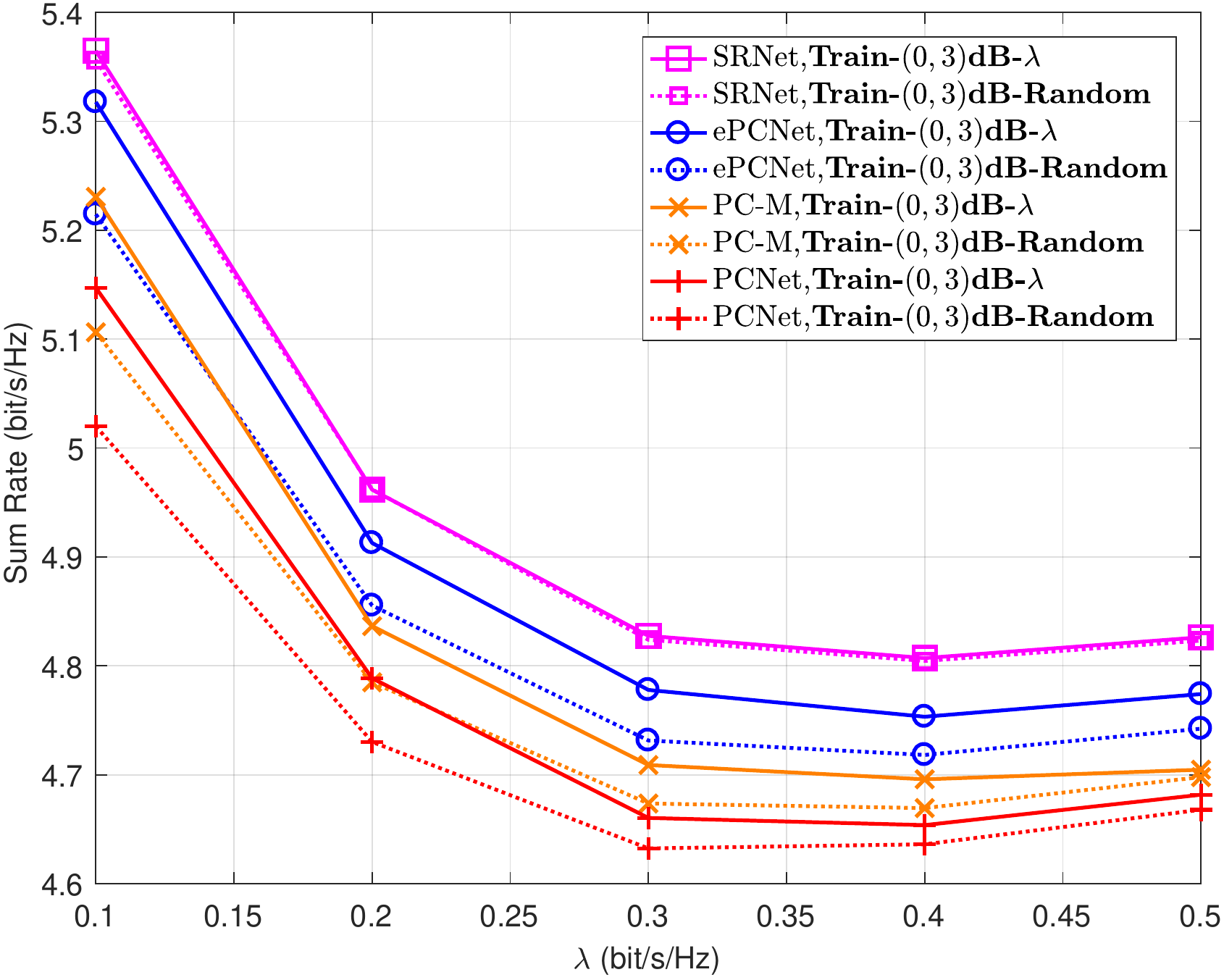}  
	\caption{Performance comparison with random rate requirements for training.}  
\label{fig-random}   
\end{figure}

An approach to alleviate the impact of model mismatch of rate requirements is to train the networks under a range of rate requirements instead of a fixed one. We evaluate the performance of this approach by employing \textbf{Train-$(0,3)$dB-Random} for training, and the results are depicted in Fig.~\ref{fig-random}. It is shown that SRNet trained under random rate requirements performs very close to the one without model mismatch, and outperforms the three existing methods. For PCNet, ePCNet, and PC-M, it can be found that the performance loss caused by model mismatch can be recovered to a certain extent, but the gap from the model-match case is still observable.

\begin{figure}
	\centering  
	\includegraphics[width=0.75\textwidth]{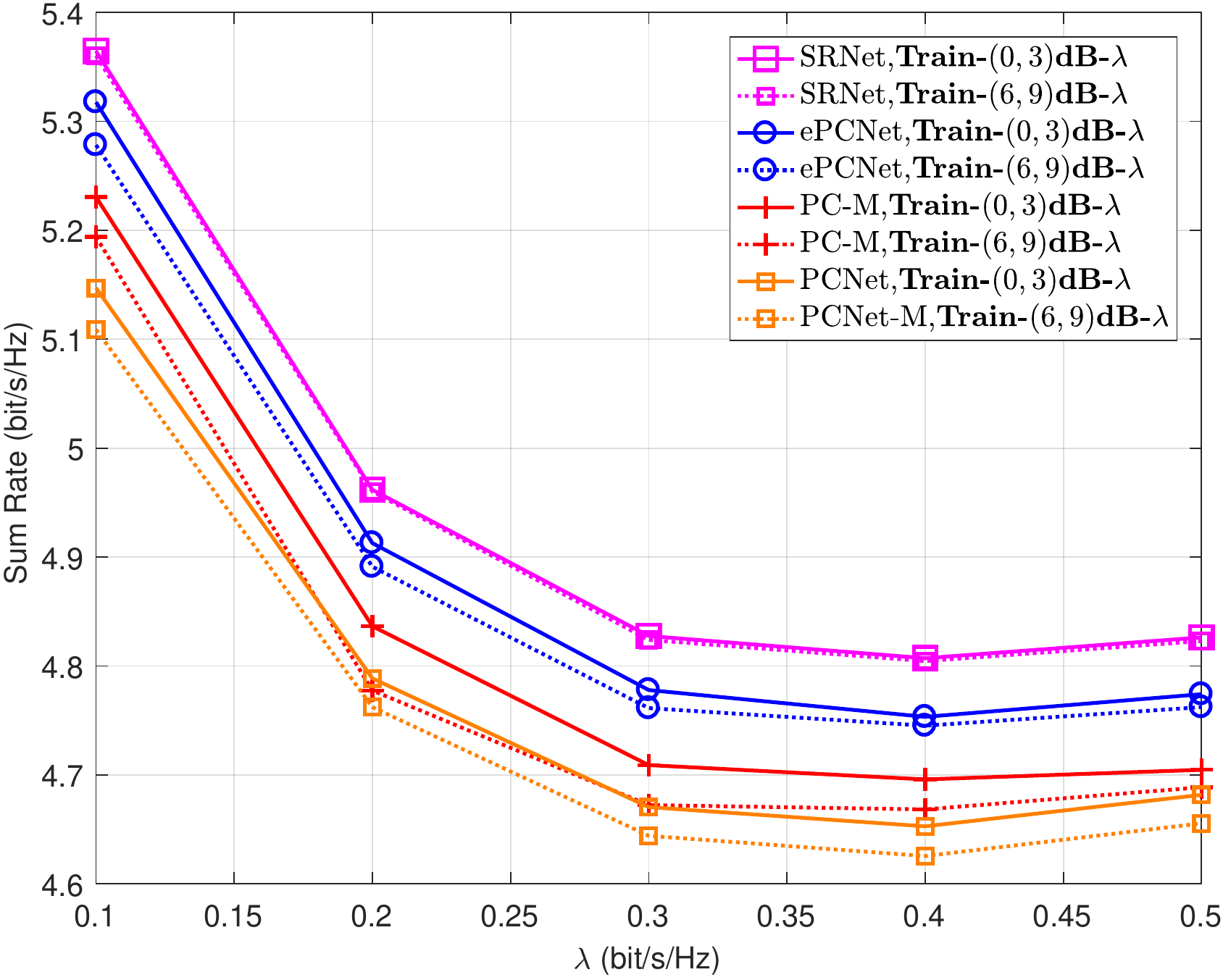}  
	\caption{Performance comparison under channel mismatch.}  
\label{fig7}   
\end{figure}

We finally examine the impact of model mismatch of channels in Fig.~\ref{fig7}. In particular, for a given testing dataset \textbf{Test-$(0,3)$dB-$\lambda$}, we train the networks with the datasets \textbf{Train-$(0,3)$dB-$\lambda$} and \textbf{Train-$(6,9)$dB-$\lambda$}, respectively, where the latter leads to model mismatch of channels. The minimum rate requirement $\lambda$ keeps identical for training and testing, and the networks are re-trained once $\lambda$ changes. It is shown that the channel mismatch has negligible impact on the performance of SRNet. Yet, the performance loss for PCNet, ePCNet, and PC-M is observable.

Based on the results in Fig.~\ref{fig-random} and Fig.~\ref{fig7}, it can be concluded that SRNet is insensitive to the model mismatch between training and testing datasets, so that we do not need to frequently re-train SRNet for different application scenarios.

\subsubsection{\underline{Complexity Comparison}}
We compare the complexity of the methods in terms of online running time. We implement the methods on a computer with Intel$^\circledR$ Core$^\text{TM}$ i7-8700K CPU (3.70GHz) without using the acceleration from GPU, and measure the average online running time for solving 10,000 SRM problems. The results are shown in Table~\ref{table2}.

We can find that, as expected, the deep learning based methods, including SRNet, SRNet-Heu, PCNet, ePCNet, and PC-M, have much shorter running time than the numerical methods requiring iterative algorithms to solve optimization problems. By using the heuristic method of selecting point $C$, the complexity of SRNet is further reduced though not significant in the three-cell setup.
The good performance and low complexity make the proposed SRNet attractive for practical applications.
%

\begin{table}
  \centering
  \caption{Average running time for solving 10,000 SRM problems.}
	\begin{tabular}{cccccccc}
	\toprule  
	 SRNet&SRNet-Heu&PCNet&ePCNet&PC-M&SQP&Interior Point \\
	\midrule  
	 184~ms & 180~ms& 120~ms& 550~ms &194~ms & 34~s&194~s\\
	\bottomrule 
\end{tabular}
\label{table2}
\end{table}

\section{Conclusions}
\label{s:6}
In this paper we proposed a deep learning based method for multicell power control to maximize the sum rate subject to per-user rate constraints and per-BS power constraints. To ensure the learned power control results to satisfy the constraints, a novel projection block was designed as a component of the DNN, which has low complexity for online implementation. Explicit-form expression of the backpropagated gradient was derived for the proposed projection block, which enables an unsupervised learning of the power control to directly maximize the sum rate. We have also developed a heuristic method to further reduce the network size of the designed DNN. Simulation results showed that the proposed method can achieve better performance with low complexity compared to existing deep learning methods. Moreover, the results showed that the proposed method is insensitive to the model mismatch between training and testing datasets, and thus it is not necessary to frequently re-train the DNN for different application~scenarios.

\appendices
\numberwithin{equation}{section}

\section{Proof of Proposition 1}
\label{a:1}

The impact of $\mathbf{d}$ on $\mathbf{p}^C$ can be observed from \eqref{E:parallel}. Upon substituting $q_i = \gamma_{i,\text{min}}\sigma^2$, we can rewrite \eqref{E:parallel} as
  \begin{align}
    \mathbf{b}_i^T\mathbf{p} =  \gamma_{i,\text{min}}\left(\sigma^2 + \frac{\|\mathbf{b}_i\|_2d_i}{\gamma_{i,\text{min}}}\right) , \forall i.
  \end{align}
It is shown that increasing $d_i$ is equivalent to increase the noise power from $\sigma^2$ to $\sigma^2 + \frac{\|\mathbf{b}_i\|_2d_i}{\gamma_{i,\text{min}}}$. Considering that the increased noise power leads to the increase of transmit powers to achieve the required minimum SINR, Proposition 1 is proved.

\section{Derivations of $\frac{\partial \mathbf{p}^{D}}{ \partial  \hat{\mathbf{p}}}$ and $\frac{\partial \mathbf{p}^{D}}{ \partial  \mathbf{p}_{C}}$}
\label{S:derivation}
Let us first derive $\frac{\partial \mathbf{p}^{D}}{ \partial  \hat{\mathbf{p}}}$.
By substituting \eqref{E:epsilon-max} into \eqref{E:pointD}, we can obtain
\begin{align} \label{E:pointD-power}
    \mathbf{p}^D = \hat{\mathbf{p}} + \frac{[\mathbf{q}-\mathbf{B} \hat{\mathbf{p}}]_{k}}{\left[\mathbf{B}\left(\mathbf{p}^{C}-\hat{\mathbf{p}}\right)\right]_{k}}(\mathbf{p}^C-\hat{\mathbf{p}}).
\end{align}

To facilitate the derivation, we first rewritten $\mathbf{p}^{D}$ given by \eqref{E:pointD-power} as
\begin{align}
\mathbf{p}^{D}&=(\hat{\mathbf{p}}-\mathbf{p}^{C})-\frac{[\mathbf{q}-\mathbf{B} \hat{\mathbf{p}}]_{k}}{\left[\mathbf{B}\left(\mathbf{p}^{C}-\hat{\mathbf{p}}\right)\right]_{k}}\left(\hat{\mathbf{p}}-\mathbf{p}^{C}\right)+\mathbf{p}^{C}\notag \\
 &=\frac{[\mathbf{B} \mathbf{p}^{C}-\mathbf{q}]_{k}}{\left[\mathbf{B}\left(\mathbf{p}^{C}-\hat{\mathbf{p}}\right)\right]_{k}}\left(\hat{\mathbf{p}}-\mathbf{p}^{C}\right)+\mathbf{p}^{C}.
\end{align}
Then, $\frac{\partial \mathbf{p}^{D}}{ \partial  \hat{\mathbf{p}}}$ can be obtained as
  \begin{small}
    \begin{align}\label{E:Appe-4}
\frac{\partial \mathbf{p}^{D}}{\partial\hat{\mathbf{p}}}
&=[\mathbf{B p}^{C}-\mathbf{q}]_{k}\left(\frac{1}{[\mathbf{B}(\mathbf{p}^{C}-\hat{\mathbf{p}})]_{k}} \frac{\partial(\hat{\mathbf{p}}-\mathbf{p}^{C})}{\partial \hat{\mathbf{p}}}+\left[\frac{\partial}{\partial\hat{\mathbf{p}}}\left(\frac{1}
{[\mathbf{B}(\mathbf{p}^{C}-\hat{\mathbf{p}})]_{k}}\right)\right] (\hat{\mathbf{p}}-\mathbf{p}^{C}\right)^{T})\notag\\
&=[\mathbf{B p}^{C}-\mathbf{q}]_{k}\left(\frac{1}{[\mathbf{B}(\mathbf{p}^{C}-\hat{\mathbf{p}})]_{k}} \mathbf{I}+\left[\frac{\partial}{\partial \widehat{\mathbf{p}}}\left(\frac{1}{\sum_{j=1}^{K}([\mathbf{p}^{C}]_{j}-
[\widehat{\mathbf{p}}]_{j})[\mathbf{B}]_{k j}}\right)\right] (\widehat{\mathbf{p}}-\mathbf{p}^{C})^{T}\right),
  \end{align}
    \end{small}
where the $i$-th element of the term $\frac{\partial}{\partial \widehat{\mathbf{p}}}\left(\frac{1}{\sum_{j=1}^{K}([\mathbf{p}^{C}]_{j}-
[\widehat{\mathbf{p}}]_{j})[\mathbf{B}]_{k j}}\right)$ can be derived as
  \begin{small}
 \begin{align}  \label{E:Appe-5}
  \frac{\partial}{\partial\left[\hat{\mathbf{p}}\right]_i}\left(\frac{1}{\sum_{j=1}^{K}([\mathbf{p}^{C}]_{j}-[\hat{\mathbf{p}}]_{j})[\mathbf{B}]_{k j}}\right)
=\frac{[\mathbf{B}]_{k i}}{\left(\sum_{j=1}^{K}([\mathbf{p}^{C}]_{j}-[\hat{\mathbf{p}}]_{j}) [\mathbf{B}]_{k j}\right)^{2}}
 =\frac{[\mathbf{B}]_{k i}}{\left\|[\mathbf{B}(\mathbf{p}^{C}-\hat{\mathbf{p}})]_{k}\right\|^{2}}.
  \end{align}
      \end{small}
With \eqref{E:Appe-4} and \eqref{E:Appe-5}, we can obtain the result of  $\frac{\partial \mathbf{p}^{D}}{ \partial  \hat{\mathbf{p}}}$ as shown by \eqref{E:BackDP}.

We next derive $\frac{\partial \mathbf{p}^{D}}{ \partial  \mathbf{p}_{C}}$.
From \eqref{E:pointD-power}, we can obtain that
\begin{small}
  \begin{align}\label{E:Appe-1}
\frac{\partial \mathbf{p}^{D}}{\partial \mathbf{p}^{C}}
&=\left[\mathbf{q}-\mathbf{B} \hat{\mathbf{p}}\right]_{k}\left(\frac{1}{[\mathbf{B}(\mathbf{p}^{C}-\hat{\mathbf{p}})]_{k}} \frac{\partial(\mathbf{p}^{C}-\hat{\mathbf{p}})}{\partial \mathbf{p}^{C}}+\left[\frac{\partial}{\partial \mathbf{p}^{C}}\left(\frac{1}{[\mathbf{B}(\mathbf{p}^{C}-\hat{\mathbf{p}})]_{k}}\right)\right] (\mathbf{p}^{C}-\hat{\mathbf{p}})^{T}\right)\notag\\
&=[\mathbf{q}-\mathbf{B} \hat{\mathbf{p}}]_{k}\left(\frac{1}{[\mathbf{B}(\mathbf{p}^{C}-\hat{\mathbf{p}})]_{k}} \mathbf{I}+\left[\frac{\partial}{\partial \mathbf{p}^{C}}\left(\frac{1}{\sum_{j=1}^{K}([\mathbf{p}^{C}]_{j}-[\hat{\mathbf{p}}]_{j})[\mathbf{B}]_{k j}}\right)\right] (\mathbf{p}^{C}-\hat{\mathbf{p}})^{T}\right),
  \end{align}
  \end{small}
where the $i$-th element of the term $\frac{\partial}{\partial \mathbf{p}^{C}} \left(\frac{1}{\sum_{j=1}^{K}([\mathbf{p}^{C}]_{j}-[\hat{\mathbf{p}}]_{j})[\mathbf{B}]_{k j}}\right)$ can be derived as
    \begin{small}
  \begin{align}  \label{E:Appe-2}
  \frac{\partial}{\partial\left[\mathbf{p}^{C}\right]_{i}}\left(\frac{1}{\sum_{j=1}^{K}(\left[\mathbf{p}^{C}\right]_{j}-[\hat{\mathbf{p}}]_{j})[\mathbf{B}]_{k j}}\right)
=\frac{-[\mathbf{B}]_{k i}}{\left(\sum_{j=1}^{K}(\left[\mathbf{p}^{C}\right]_{j}-[\widehat{\mathbf{p}}]_{j}) [\mathbf{B}]_{k j}\right)^{2}}
  &=\frac{-[\mathbf{B}]_{k i}}{\left\|\left[\mathbf{B}(\mathbf{p}^{C}-\hat{\mathbf{p}})\right]_{k}\right\|^{2}}.
  \end{align}
  \end{small}
With \eqref{E:Appe-1} and \eqref{E:Appe-2}, we can obtain the result of $\frac{\partial \mathbf{p}^{D}}{ \partial  \mathbf{p}_{C}}$ as shown by \eqref{E:BackDC}.

\section{Solution to Problem \eqref{E:problem-appro}}\label{a:2}
We prove the solution by contradiction.
Suppose that $d_{\max}^\star\mathbf{1}$ is not the optimal solution, and let $\mathbf{d}=[d_1,\dots,d_K]^T$ denote the optimal solution, which needs to satisfy $\min\{\mathbf{d}\} > d_{\max}^\star$, i.e., $d_{\max}^\star\mathbf{1}\preceq\mathbf{d}$. Then, from Proposition 1, we know that $\mathbf{p}^C(d_{\max}^\star\mathbf{1})\preceq\mathbf{p}^C(\mathbf{d})$, which leads to $\min\{P_{\text{max}}\mathbf{1}-\mathbf{p}^C( d_{\max}^\star\mathbf{1})\}> \min\{P_{\text{max}}\mathbf{1}-\mathbf{p}^C(\mathbf{d})\}$.
From \eqref{E:dmax}, we know that $\min\{P_{\text{max}}\mathbf{1}-\mathbf{p}^C( d_{\max}^\star\mathbf{1})\}=0$, which means that $\min\{\mathbf{p}^C(\mathbf{d})\}>P_{\text{max}}$, violating constraint \eqref{E:problem-appro}. Thus, such a solution $\mathbf{d}$ does not exist, and $d_{\max}^\star\mathbf{1}$ is the optimal solution.

%


\section*{Acknowledgement}
The authors would like to thank Dr. Fei Liang for sharing the code of PCNet.

\bibliographystyle{IEEEtran}
\bibliography{IEEEabrv,ref}
\end{document}